\documentclass[12pt,a4paper]{article}

\usepackage{amsmath} 
\usepackage{amssymb}
\usepackage{graphicx}

\begin{document}
\title{Bubble wall correlations in cosmological phase transitions}

\author{
Ariel M\'{e}gevand\thanks{Member of CONICET, Argentina. E-mail address:
megevand@mdp.edu.ar}~ 
and Federico Agust\'{\i}n Membiela\thanks{Member of CONICET, Argentina. E-mail
address: membiela@mdp.edu.ar} \\[0.5cm]
\normalsize \it IFIMAR (CONICET-UNMdP)\\
\normalsize \it Departamento de F\'{\i}sica, Facultad de Ciencias Exactas
y Naturales, \\
\normalsize \it UNMdP, De\'{a}n Funes 3350, (7600) Mar del Plata, Argentina }
\date{}
\maketitle

\begin{abstract}
We study statistical relationships between bubble walls in cosmological first-order
phase transitions. We consider the conditional
and joint probabilities  
for different points on the walls to remain uncollided at given times. 
We use these results to discuss space and time correlations of bubble walls
and their relevance for the consequences of the transition.
In our statistical treatment, the kinematics of bubble nucleation and growth is characterized 
by the nucleation rate and the wall velocity as functions of time. We
obtain general expressions in terms of these two quantities, and 
we consider several specific examples and applications.
\end{abstract}

\section{Introduction}

\label{intro}

It is well known that first-order phase transitions may have occurred 
in the early universe, and may have left several potentially observable remnants.
In a cosmological first-order phase transition, a metastable high-temperature phase (false vacuum) 
undergoes supercooling, 
and then
the phase transition proceeds through the nucleation and expansion of 
bubbles of the low-temperature stable phase (true vacuum). 
The dynamics is different in the case of a 
``vacuum'' transition and
in the case of a ``thermal'' transition \cite{tww92,kkt94}.
In the former case, the nucleation of bubbles occurs in the absence of a plasma, and
the nucleation rate $\Gamma$ is given by the probability of decay of the false vacuum per unit time per unit volume
\cite{c77,cc77}. Besides,
all the false-vacuum energy, which is released at the bubble
walls, goes into accelerating the latter, which reach velocities $v\simeq 1$. 
This may also occur in a thermal phase transition with extreme supercooling, 
in which the wall velocity may exhibit runaway behavior \cite{bm09,bm17,hklt20}.
The bubble walls disappear as bubbles collide, and the energy stored in the walls is transferred
to thermal energy. 
On the other hand, a thermal transition occurs
in the presence of a plasma, and we have a temperature-dependent nucleation rate $\Gamma(T)$ \cite{l81,l83}.
In this case,
the walls generally reach a terminal velocity $v(T)$, and  most
of the released energy (latent heat) goes to the fluid. 
Therefore, as the walls move, a reheating of the plasma occurs, as well as bulk
fluid motions.

Even in the thin wall approximation, which is generally valid, the dynamics of 
thermal phase transitions is complex. In the first place, 
the nucleation rate is very sensitive to temperature variations.
In the second place, the wall velocity depends on the microphysics which determines 
the friction with the plasma \cite{t92,mp95}, and is also affected by the hydrodynamics \cite{w84,l94,ikkl94}.
Nevertheless, in many cases it is possible to assume 
that the nucleation rate is homogeneous 
and that the bubbles are spherical and all expand with the same speed.
In such cases, the bubble kinematics is determined by the two basic ingredients $\Gamma(t)$ and $v(t)$.
The kinematics is also affected, to a greater or lesser
extent, by the scale factor $a(t)$. In most cases, however, the variation
of the latter can be neglected for the duration of the phase
transition. An exception is the case of strongly-supercooled phase transitions \cite{tww92,mr17,eln19}.
In the statistical treatment of the phase transition, the quantities are averaged over possible realizations, and 
we shall denote $\langle Q\rangle$ the ensemble average of a quantity $Q$.
In practice, these averages are calculated from the average number of bubbles nucleated in a given volume during a certain time, which is given by $\Gamma(t)$.

In the development of the transition, 
the most evident measure of progress is the volume fraction occupied by bubbles, 
$f_{b}(t)$.
However, other quantities can 
be used as well, such as the fraction of bubble wall which remains uncollided,  $f_{S}(t)$  
\cite{tww92}. Since the collided walls 
quickly\footnote{In general, all the terms in the equation of motion for the scalar field (order parameter)
involve a single scale, namely, the scale of the theory, which is $\sim T$, 
so the characteristic time scale for the wall dynamics is $\sim T^{-1}$. This time is generally much shorter than the 
duration of the phase transition, which is determined by the Hubble rate and involves also the Planck scale. \label{notaTt}} 
disappear inside merged bubbles\footnote{For recent discussions on the behavior of the 
scalar field after bubble collisions, see \cite{jkt19} and references therein.}, 
the uncollided wall area is essentially the total area $S_\mathrm{tot}$ that is physically present at time $t$. 
We have
$S_{\mathrm{tot}}=\sum_i S_i$, where $S_i$ is the wall area of bubble $i$ which remains 
(uncollided) at time $t$. 
The fraction of surface  $f_{S}$ is defined as $S_\mathrm{tot}$ divided by the total area $\sum_i 4\pi R_i^2$ of bubbles of radii $R_i$, 
including area that has been covered by bubbles.
This quantity varies from $f_{S}=1$ at the beginning of
the phase transition (when bubbles are isolated) to $f_{S}=0$ at
the end (when all bubbles have merged and their walls have disappeared).
The quantity $f_{S}$ 
tracks the conversion of potential energy (false vacuum energy or latent heat)
to other forms of energy (kinetic energy of the wall, kinetic energy of the fluid, or thermal energy).
It  will be more relevant than $f_{b}$ to those
processes involving the bubble walls. In particular, the 
departures from equilibrium which give rise to the important consequences of the phase transition 
originate at the bubble walls.
Let us consider a few examples.

\paragraph{Baryogenesis.}

If the electroweak phase transition is of first order, the walls of expanding
bubbles push a net charge density into the symmetric phase, which
bias baryon-number violating processes \cite{krs85,ckn93,rt99,mr12}.
This mechanism relies on diffusion processes that take place up to
a distance $l$ from the wall, which is naturally $l\sim T^{-1}$.
This length is several orders of magnitude smaller than the typical
bubble radius $R$, which is of order $H^{-1}$, where $H$ is the
Hubble rate. It is in this very thin shell next to the bubble walls
where baryon number generation occurs. 
The baryon number density $n_B(t)$ which is left behind by the walls depends on the value
of the wall velocity. 
The latter is often estimated at
the onset of nucleation, although $v(t)$ generally varies during the phase
transition.

\paragraph{Gravitational waves.}

The energy that is set in motion by the bubble walls is a source of gravitational waves (GWs) \cite{tw90,gwlisa,gwlisa2}.
This energy may be concentrated in the walls themselves, or it may
be transferred to bulk fluid motions \cite{ktw92,kkt94}. 
In the latter case, the relation of the wall surface with the generated GW spectrum 
is indirect, since the direct source of GWs is the turbulence \cite{kkt94} 
or the sound waves \cite{hhrw14} caused in the fluid, which may last longer than the phase transition. 
In any case, the walls are 
the source of such fluid motions.
Furthermore, in the so called bubble-collision
mechanism the energy transferred to the fluid is assumed to be concentrated
in a thin shell around the walls. 
In the envelope approximation \cite{kt93}, 
the contribution of the overlap regions to the gravitational radiation is neglected, and
the energy-momentum tensor is concentrated in the ``envelope''
of walls surrounding a cluster of bubbles.

\paragraph{Topological defects.}

Perhaps the simplest example illustrating the formation of topological
defects \cite{k76} is the trapping of a vortex in two spatial dimensions.
Consider the spontaneous symmetry-breaking of a global $U(1)$ symmetry,
where a complex scalar field $\phi$ vanishes in the symmetric phase
and takes nonvanishing values $\phi=ve^{i\alpha}$ in the broken-symmetry
phase. The modulus $v$ is fixed but the phase $\alpha$ is arbitrary
and is uncorrelated in different bubbles. When two bubbles meet, $\alpha$
will rearrange itself so that it varies smoothly from one bubble to
the other. Moreover, this phase will tend to take a constant value
throughout space. When three bubbles meet at a given point, $\alpha$
will tend to vary smoothly along a closed line across the three bubbles.
However, a complete equilibration to reach a constant phase may be
topologically impossible, in which case a defect will be trapped in
a symmetric-phase region enclosed by the bubbles. Nevertheless,
it is very unlikely that three bubble walls collide simultaneously
at a single point. Two of the bubbles will meet first, and the third
one will arrive later. If the phase equilibration \cite{kv95} between
the first two bubbles completes before the arrival of the third bubble,
the formation of a vortex may be avoided \cite{bkvv95}.

\bigskip

It is clear that the wall dynamics plays a relevant role in the determination
of these consequences of the phase transition. In particular, electroweak
baryogenesis takes place in a thin shell next to the bubble walls.
Since the baryon number density depends on the wall velocity, 
a precise evaluation of the baryon asymmetry requires integrating 
$n_B(t)$ in time, weighted with the volume 
$\langle S_\mathrm{tot}(t)\rangle v(t)dt$. 
In the case of GW production (at least in the envelope approximation), 
the energy-momentum tensor $T$ is also
localized in a thin shell around the bubble walls. However, the spectrum
of GWs depends on the correlation function $\langle T(x)T(y)\rangle$
between different space-time points (see, e.g., \cite{cds08,jt17}). One would then
expect that the GW spectrum will be related to bubble surface correlations 
$\langle S(t)S(t')\rangle$ 
rather than to the average area. On the other hand,
for topological defect formation, the probability of trapping a defect
in a sequence of bubble collisions depends on the probability
that a point on a given bubble wall will soon collide once a nearby point
has already collided. This mechanism should then be related to the
correlation between different points on the same bubble wall.

It is well known that 
the probability that a random point of space is in the false vacuum
is the same as
the fraction of volume remaining in that phase. Also, the fraction of the bubble
wall that remains uncollided is given by the probability that a point
\emph{on a bubble wall} remains in the false vacuum (this probability
is not the same as the previous one, since the nucleation in the vicinity
of the point is affected by the presence of the bubble to which it
belongs \cite{tww92}).
Similarly, the surface correlations mentioned above will depend on
conditional or joint probabilities for multiple points belonging to
bubble walls.

In the present paper, we consider such probabilities. In the next section
we review some existing results and discuss the probability that a
set of arbitrary points in space remain in the false vacuum at a given
time. In Sec.~\ref{pointsonbubbles} we calculate the probability
that a point in the surface of a bubble is uncollided, depending on
whether another point in the same bubble wall or in the wall of another
bubble is still uncollided. In section \ref{somecases} we apply these results
to discuss 
space and time correlations in the envelope of uncollided walls.
We compare the results for different models.
In Sec.~\ref{cosmo} we consider some specific applications to the cosmological
consequences of the transition.
Finally, in Sec.~\ref{conclu}, we summarize our conclusions.

\section{Phase transition dynamics and probability of remaining in the false
vacuum}

\label{ptdyn}

\subsection{Global dynamics}

As already mentioned, the relevant quantities in the
kinematics of the phase transition are the nucleation rate $\Gamma$ and the wall velocity $v$.
We shall assume that these quantities are homogeneous and given by certain functions of time 
$\Gamma(t)$, $v(t)$. In particular, a homogeneous $v(t)$ implies that bubbles are spherical and 
that all expand with the same velocity.
Although these assumptions are very common, they are not always valid.
The nucleation rate is certainly homogeneous
in the case of a vacuum transition, where $\Gamma$ is of the
form $\Gamma=Ae^{-S}$, with $A$ and $S$ constant \cite{c77,cc77}.
Furthermore, in this case the wall velocity quickly approaches the asymptotic value $v=1$. 
In the case of a thermal transition, $\Gamma$ has a similar
form, but $A$ and $S$ depend on the temperature \cite{l81,l83}, and may thus have 
space and time variations.
In general, the
bubble walls reach a terminal velocity in a time which is
much shorter than the total duration of the phase transition, but this velocity 
depends also on the temperature. 
In a phase transition mediated
by detonations \cite{s82}, the latent heat that is released at the bubble walls
only reheats the plasma behind the walls (inside the bubbles), so $\Gamma$ and $v$  are not affected by temperature inhomogeneities.
In contrast, for deflagrations \cite{gkkm84,k85,kl95}, the fluid is perturbed in front of the bubble walls, and
perturbations coming from different bubbles cause inhomogeneous reheating \cite{ikkl94}.
Even in this case, homogeneous $\Gamma$ and $v$ are often assumed to study the 
kinetics of the phase transition (see, e.g., \cite{eikr92}).
Nevertheless, in the case of very slow deflagrations (with $v\lesssim 0.1$), 
the released latent
heat is distributed with a relatively high speed, so a homogeneous
reheating can be assumed  \cite{h95}.

In this section and the following we shall consider  arbitrary 
functions $\Gamma(t)$ and $v(t)$, while
in Secs.~\ref{somecases} and \ref{cosmo} we shall consider particular examples.
Actually, rather than the wall velocity, the basic ingredient here will be
the bubble radius. Between two times $t'$ and $t$, the
radius of a bubble increases by 
\begin{equation}
R(t',t)=\int_{t'}^{t}v(t'')dt''.\label{radio}
\end{equation}
For simplicity, we shall ignore the effect of the scale factor $a(t)$,
which would introduce a factor $a(t)/a(t'')$ in the integrand. As already mentioned, this approximation
is valid in most cases. In any case, generalizing our treatment to include this
effect should be straightforward. If we neglect the initial radius
of the bubble at the time of its nucleation, which is also a good
approximation in general, then Eq.~(\ref{radio}) gives  the
radius of a bubble which has nucleated at time $t'$ and has expanded until
time $t$.
The fraction of volume occupied by bubbles is given by $f_{b}=1-P_{fv}$,
where $P_{fv}$ is the fraction of volume in the false
vacuum, which coincides with the probability that an arbitrary point
is in that phase. This quantity is well known
\cite{gt80}. We shall consider a derivation here, which we shall
generalize to less simple cases below.

\subsection{Probability that a given point in space remains in the false vacuum}

\label{frac}

By time $t$, a point $p$ may have been reached by bubbles nucleated
at times $t''<t$. We begin by considering the probability $P_{\mathrm{out}}(t'')$
that $p$, at time $t$, is outside of any bubbles nucleated before a certain
$t''$. 
This probability depends also on $t$, which we omit for simplicity of notation. 
Then, the probability that $p$ remains outside of any bubbles
nucleated before $t''+dt''$ is given by the product 
\begin{equation}
P_{\mathrm{out}}(t''+dt'')=P_{\mathrm{out}}(t'')(1-dP(t'')),\label{eqPout}
\end{equation}
 where the last factor is the probability that $p$ was not reached
by bubbles nucleated between $t''$ and $t''+dt''$ either. That is to say,
$dP(t'')$ is the probability that $p$ has been reached by some bubble
nucleated between $t''$ and $t''+dt''$, assuming that $p$ was not
reached by bubbles nucleated before $t''$. From Eq.~(\ref{eqPout})
we readily obtain a differential equation whose solution is 
\begin{equation}
P_{\mathrm{out}}(t'')=e^{-\int_{t_{c}}^{t''}dP}.\label{solPout}
\end{equation}
 Here, $t_{c}$ is the initial time, corresponding to the critical
temperature $T_{c}$ of the phase transition, before  which the nucleation
rate vanishes. Evaluating at $t''=t$ we obtain the probability that
the point $p$ remains in the false vacuum at time $t$,
\begin{equation}
P_{fv}(t)=e^{-\int_{t_{c}}^{t}dP(t'')}.\label{Pfvini}
\end{equation}

\begin{figure}[tbh]
\centering
\includegraphics[height=5cm]{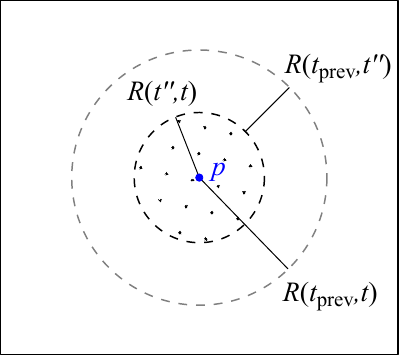}
\caption{The region in which bubbles must nucleate at time $t''$ in order
to reach the point $p$ before time $t$ (dots). The outer circle
indicates the region within which bubbles should nucleate at $t_{\mathrm{prev}}<t''$
in order to affect the dotted region. 
\label{figfrac}}
\end{figure}

We still have to compute the conditional probability $dP(t'')$ that
(at time $t$) $p$ is inside a bubble nucleated between $t''$ and
$t''+dt''$, assuming that it is outside of any previously nucleated
bubbles. For a bubble nucleated at time $t''$ to reach the point
$p$ before time $t$, the bubble must have nucleated at a distance
smaller than $R(t'',t)$ from the point. In Fig.~\ref{figfrac},
the dots represent the possible nucleation points. To calculate the
probability that a bubble was nucleated within this radius at time
$t''$, we must determine whether, at that time, the whole region
was actually available for bubble nucleation, since part of the space
could have been occupied by previously nucleated bubbles. 
Nevertheless, such bubbles would also reach the point $p$ before time $t$,
which we are assuming does not occur.
Indeed, consider a bubble
nucleated at a certain $t_{\mathrm{prev}}<t''$. For this bubble to
invade the dotted region at $t''$, it must have nucleated at a distance
smaller than $R(t_{\mathrm{prev}},t'')$ from it (see Fig.~\ref{figfrac}).
But then it would be too close to $p$, at a distance smaller than
$R(t_{\mathrm{prev}},t'')+R(t'',t)=R(t_{\mathrm{prev}},t)$. 

Hence,  the whole volume of the dotted region 
is free of bubbles at time $t''$ and 
is available for nucleations. Thus, the probability $dP(t'')$ is given by 
\begin{equation}
dP(t'')=dt''\Gamma(t'')\frac{4\pi}{3}R(t'',t)^{3}.\label{dPfv}
\end{equation}
 From (\ref{Pfvini}-\ref{dPfv}) we obtain 
\begin{align}
P_{fv}(t)=e^{-I(t)},\label{Pfv}
\end{align}
where 
\begin{equation}
I(t)=\int_{t_{c}}^{t}dt''\Gamma(t'')\frac{4\pi}{3}R(t'',t)^{3}.\label{I}
\end{equation}

\subsection{Probability that a point is in the false vacuum given that another
point is in the false vacuum}

\label{fv2}

Let us now consider the probability $P_{p|p'}$ that a point $p$
remains in the false vacuum at time $t$, given that another point
$p'$ was in the false vacuum at time $t'\leq t$. Proceeding as before, we consider
the probability $P_{\mathrm{out}}(t'')$ that $p$ has
not been reached by bubbles nucleated before $t''$, and then the probability
$dP(t'')$ that $p$ has been reached by a bubble nucleated between
$t''$ and $t''+dt''$. Thus, we obtain the same equation for $P_{\mathrm{out}}(t'')$,
Eq.~(\ref{eqPout}), which leads to 
Eqs.~(\ref{solPout}) and (\ref{Pfvini}).
Like in the previous case,  $dP(t'')$ is
the conditional probability 
that $p$ is inside a bubble nucleated between $t''$ and $t''+dt''$
subjected to the condition that
$p$ is outside of bubbles nucleated before $t''$.
The difference is that, 
in the present case, we also have the condition that
the other point, $p'$, is in the false vacuum at time $t'$.
Therefore, we write 
\begin{equation}
P_{p|p'}=e^{-\int_{t_{c}}^{t}dP(t'')},\label{Pppini}
\end{equation}
and we must re-evaluate the conditional probability $dP(t'')$. 

\begin{figure}[tbh]
\centering
\includegraphics[height=5cm]{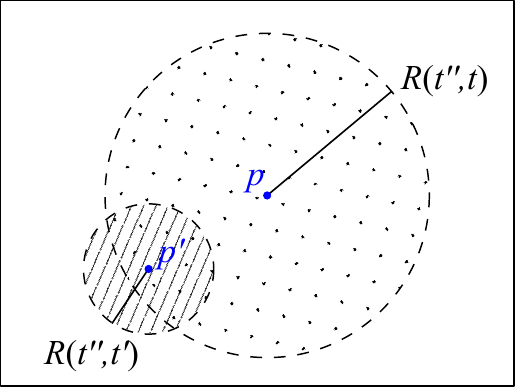}
\caption{Regions affecting the points $p$ and $p'$ for $t''<t'<t$. The dotted
region is that in which a bubble must nucleate at time $t''$ in order
to reach the point $p$ before time $t$. If the nucleation occurs
inside the striped region, the bubble would eat the point $p'$ before
time $t'$.}
\label{figfv2}
\end{figure}

At
$t''$, the bubble affecting $p$ must have nucleated within a sphere
of radius $R(t'',t)$ centered at this point, like in Fig.~\ref{figfrac}. 
Then, in principle,
we would obtain Eq.~(\ref{dPfv}). Again,
under the present conditions the dotted region is not affected, at
time $t''$, by previously nucleated bubbles. However, the nucleations
at $t''$ might reach the point $p'$ before time $t'$, which is now
forbidden by the conditional probability. For $t''>t'$ this will
not happen, so we still have 
\begin{equation}
dP(t'')=dt''\Gamma(t'')\frac{4\pi}{3}R(t'',t)^{3}\qquad(t''>t').\label{dPpfv}
\end{equation}
But for $t''\leq t'$, any nucleation at time $t''$ must occur at
a distance larger than $R(t'',t')$ from $p'$ in order to avoid affecting
this point. This situation is represented in Fig.~\ref{figfv2}.
A nucleation at $t''$ must occur inside the dotted region in order
to affect the point $p$ but outside the striped region to leave
$p'$ unaffected. Therefore, we have 
\begin{equation}
dP(t'')=dt''\Gamma(t'')\left[\frac{4\pi}{3}R(t'',t)^{3}-V_{\cap}\right]\qquad(t''\leq t'),\label{dP2}
\end{equation}
where $V_{\cap}$ is the volume of the intersection of the two spheres.

From Eqs.~(\ref{Pppini}-\ref{dP2}), we obtain 
\begin{equation}
P_{p|p'}=\exp\left[-\int_{t_{c}}^{t}dt''\Gamma(t'')\frac{4\pi}{3}R(t'',t)^{3}+\int_{t_{c}}^{t'}dt''\Gamma(t'')V_{\cap}\right].\label{Pcond0}
\end{equation}
The intersection volume $V_\cap$ depends on the radii 
\begin{equation}
r\equiv R(t'',t)\,,\quad r'\equiv R(t'',t') \label{defradsr-1}
\end{equation}
and on the separation $s$ between $p$ and $p'$.
It is given by 
\begin{equation}
V_{\cap}=\begin{cases}
4\pi r^{\prime 3}/3 & \quad \mathrm{for} \ s\leq r-r',\\
\frac{\pi}{12}(r+r'-s)^{2}\left[s+2(r+r')-\frac{3(r-r')^{2}}{s}\right] & 
\quad \mathrm{for} \ r-r'<s\leq r+r',\\
0 & \quad \mathrm{for} \ s>r+r'.
\end{cases}\label{VI}
\end{equation}
Notice that, if the separation is small enough, the smaller sphere
is completely contained inside the larger one\footnote{Remember that we are considering the specific case $t'\leq t$, so
we have $r'\leq r$. For $t'>t$ (in which case the probability $P_{p|p'}$
is conditioned to the point $p'$ being in the false vacuum in the
future) the calculation is similar, and the result is essentially
the same. To take into account this possibility, the limit of integration $t'$ in the second
integral of Eq.~(\ref{Pcond0}) must be replaced with $t_{m}=\min\{t,t'\}$. \label{notatmaytp}}; hence the value $4\pi r^{\prime 3}/3$. On the other hand, if the separation
is large enough, the intersection is empty and we have $V_{\cap}=0$
(see Fig.~\ref{figfv2}). Finally, we write Eq.~(\ref{Pcond0})
in the form 
\begin{equation}
P_{p|p'}(t,t',s)=\exp\left[-I(t)+I_{\cap}(t,t',s)\right],\label{Pcond1}
\end{equation}
where the function  $I(t)$ is given by Eq.~(\ref{I}), and we have defined the
quantity 
\begin{equation}
I_{\cap}(t,t',s)=\int_{t_{c}}^{t'}dt''\Gamma(t'')V_{\cap}(r,r',s).\label{Iint}
\end{equation}

\subsection{Probability that multiple points remain in the false vacuum}

\label{multiple}

Although we are mostly interested in points on bubble walls, we shall comment on the probability for several arbitrary points to remain in the false vacuum. 
We have obtained the probability $P_{p|p'}(t,t',s)$ of the point
$p$ being in the false vacuum at time $t$, under the condition
that $p'$ was in the false vacuum at time $t'\leq t$. Multiplying
by the probability $P_{fv}(t')$ that $p'$ was in the false vacuum
at time $t'$, Eq.~(\ref{Pfv}), we obtain the joint 
probability\footnote{For this joint probability, there is no loss of generality in the
assumption $t'\leq t$.} that $p$ is in the false vacuum at time $t$ and $p'$ is in the
false vacuum at time $t'$, 
\begin{equation}
P_{fv}^{(2)}(t,t',s)=P_{fv}(t')P_{p|p'}(t,t',s)=\exp\left[-I(t)-I(t')+I_{\cap}(t,t',s)\right] \label{Pfv2}
\end{equation}
(we denote the two-point case with a superscript 2).
The exponent in the last expression can be written as $-I_\cup$, with
\begin{equation}
I_{\cup}(t,t',s)= \int_{t_{c}}^{t}dt''\Gamma(t'')V_{\cup},
\end{equation}
where $V_{\cup}$ is the volume of the union of the two spheres of
radii $R(t'',t)$ and $R(t'',t')$,
\begin{equation}
V_{\cup}=\frac{4\pi}{3}r^{3}+\left[\frac{4\pi}{3}r^{\prime3}
+V_{\cap}(r,r',s)\right]\Theta(t'-t'').
\label{Vunion}
\end{equation}
This expression takes into account the fact that there is no sphere
of radius $R(t'',t')$ for $t''>t'$.

\begin{figure}[tbh]
\centering
\includegraphics[height=5cm]{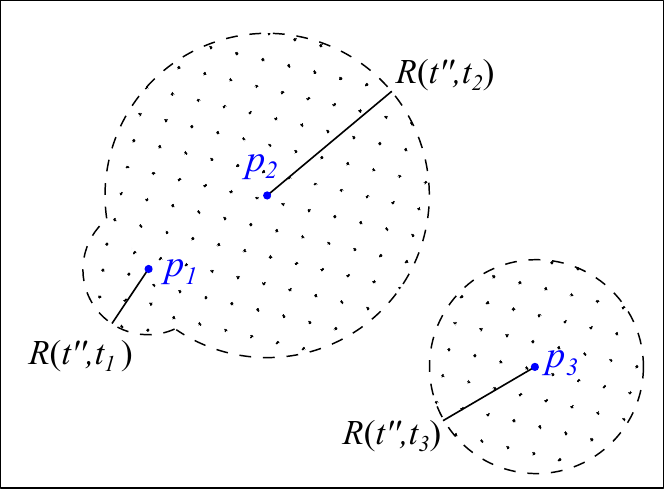}
\caption{The region in which bubbles must nucleate at time $t''$ in order
to reach at least one of the points $p_{i}$ before time $t_{i}$.
\label{figfv3}}
\end{figure}

We could have obtained this 
result as a generalization of the calculation of 
$P_{fv}^{(1)}(t)\equiv P_{fv}(t)$.
In this case, $P_{\mathrm{out}}(t'')$ would denote
the probability that \emph{none} of the two points $p,p'$ has been
eaten by bubbles nucleated before time $t''$, and $dP(t'')$ the
probability that \emph{at least one} of them has been reached by a
bubble nucleated between $t''$ and $t''+dt''$. This leads to the
total volume $V_{\cup}$.\footnote{See 
also Ref.~\cite{jt17} for an alternative calculation of $P_{fv}^{(2)}$ using past
light cones of the two events $(t,p)$, $(t',p')$. Although a 
constant velocity was assumed there, the derivation
is valid in general, and our result (\ref{Pfv2})-(\ref{Vunion}) should be equivalent to 
Eqs.~(42)-(45) of that paper. Indeed, this can be verified by comparing 
Eqs.~(46)-(47) of \cite{jt17} with our Eq.~(\ref{Iintexp})
for the specific case of an exponential 
nucleation rate which we consider in Sec.~\ref{somecases}.}
The generalization to the probability that $n$ points $p_{1},\ldots,p_{n}$
remain in the false vacuum at times $t_{1},\ldots,t_{n}$, respectively,
is straightforward. For a bubble nucleated at time $t''$ to reach
any of the points $p_{i}$ before the corresponding time $t_{i}$,
the bubble must have nucleated within one of the spheres of radius
$R(t'',t_{i})$ centered at $p_{i}$. This is illustrated in Fig.~\ref{figfv3}
for the case of three points. The result involves the volume $V_{\cup}$
of the union of the $n$ spheres, which depends on the separations
$s_{ij}$ between the different points $p_{i}$ as well as on the
radii $R(t'',t_{i})$. Thus, we have 
\begin{equation}
P_{fv}^{(n)}(t_{i},s_{ij})=e^{-I_{\cup}(t_{i},s_{ij})},\label{Pfvn}
\end{equation}
where $I_{\cup}=\int_{t_{c}}^{t_{\max}}dt''\Gamma(t'')V_{\cup}$ and
$t_{\max}=\max\{t_{i}\}$. The computation of $V_{\cup}$ must take
into account that for $t''>t_{i}$ we have $R(t'',t_{i})=0$. Care
must also be taken of avoiding over-counting the intersections, which
may be multiple. 

\section{Points on bubble walls}

\label{pointsonbubbles}

The calculations of the previous section can be adapted to points
on bubble walls. 
Considering the bubbles as overlapping spheres, 
a given point of a wall has
not collided if it has not
been eaten by another bubble. 
For joint probabilities, perhaps the
most direct approach is to consider, like in Sec.~\ref{multiple},
the whole region of bubble nucleations at time $t''$ (the dotted
region in Fig.~\ref{figfv3}). However, we are also interested in
conditional probabilities, so we shall follow the steps of Sec.~\ref{fv2}.

\subsection{Probability that a point of a bubble wall remains uncollided}

\label{fracsup}

It is instructive to consider first the simpler case of a single point,
which was first discussed in Ref.~\cite{tww92}. Since a single
bubble has a negligible contribution to the fraction of volume occupied
by bubbles, it seems, at first sight, that the probability of a given
point $p$ on its surface remaining uncollided at time $t$ will be
given by the fraction of volume $P_{fv}(t)$. However, the presence
of the reference bubble to which $p$ is attached modifies the probability
that $p$ remains in the false vacuum.

\begin{figure}[tbh]
\centering
\includegraphics[height=5cm]{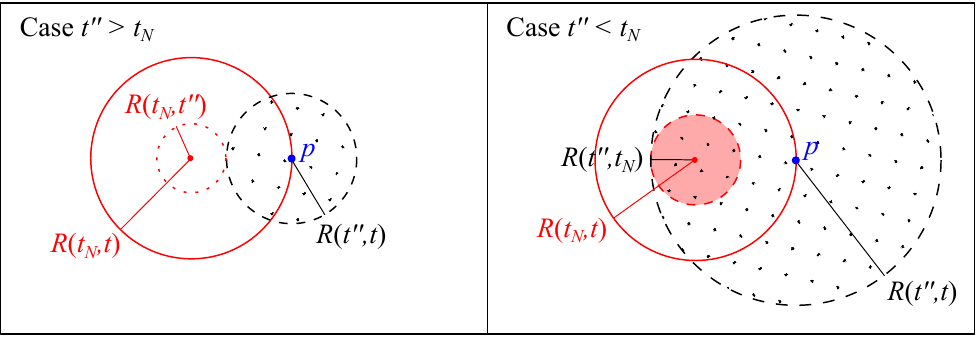}
\caption{A bubble $B$ nucleated at time $t_{N}$ (in red) whose wall
contains the point $p$, and the region where bubbles must nucleate
at time $t''$ (dots) in order to reach $p$ before time $t$. The
shaded region corresponds to nucleations at $t''$ which would prevent
the nucleation of the reference bubble. 
\label{figfracsup}}
\end{figure}

Like in Sec.~\ref{frac}, we begin by considering the probability
that the point $p$ is outside of any bubble nucleated before some time
$t''<t$. This leads to the differential equation (\ref{eqPout}) and
its solution (\ref{Pfvini}). Thus, we obtain the probability that
$p$ is uncollided,
\begin{equation}
P_{u}=e^{-\int_{t_{c}}^{t}dP(t'')},\label{Pncini}
\end{equation}
where, like before, $dP(t'')$ is the probability of $p$ being inside
a bubble nucleated between $t''$ and $t''+dt''$, assuming that it
is not inside any bubble nucleated before $t''$. Again, for this to
happen, a bubble must have nucleated at a distance smaller than $R(t'',t)$
from $p$ (the dotted region in Fig.~\ref{figfrac}). For the present
case, Fig.~\ref{figfracsup} shows the dotted region as well as the
wall which contains $p$ (represented with a solid red line). The
corresponding bubble was nucleated at a certain time $t_{N}$, and
we may have $t_{N}<t''$ or $t_{N}>t''$.

We need to determine which part of the dotted region is actually available
for bubble nucleation at time $t''$. It is straightforward to show
that, like in the previous section, the dotted region could not be
invaded at time $t''$ by bubbles nucleated at previous times\footnote{In particular, in the case $t_{N}<t''$, the reference bubble wall
containing $p$ will be, at time $t''$, just touching the limit of
the dotted region, since $R(t_{N},t'')+R(t'',t)=R(t_{N},t)$. This
is sketched with a red dotted circle in the left panel of Fig.~\ref{figfracsup}.} 
$t_{\mathrm{prev}}<t''$. On the other hand, in the case $t''<t_{N}$,
a bubble nucleated at $t''$ may prevent the nucleation of the reference
bubble $B$. This will happen if the former nucleates too close to the
nucleation point of the latter; specifically, within a radius $R(t'',t_{N})$
(shaded region in the the right panel of Fig.~\ref{figfracsup}).
Since we are assuming that bubble $B$ exists, no bubbles
can have nucleated in this region at time $t''$. The probability
that a bubble nucleates in the remaining part of the dotted region
at a time between $t''$ and $t''+dt''$ is given by\footnote{The forbidden (shaded) region is always completely contained inside
the sphere with dots, since $R(t'',t_{N})+R(t_{N},t)=R(t'',t)$.} 
\begin{equation}
dP(t'')=dt''\Gamma(t'')\frac{4\pi}{3}\left[R(t'',t)^{3}-R(t'',t_{N})^{3}\right]\label{dpant}
\end{equation}
 for $t''<t_{N}$. In contrast, for $t''>t_{N}$, the whole dotted
region is available, and we have
\begin{equation}
dP(t'')=dt''\Gamma(t'')\frac{4\pi}{3}R(t'',t)^{3}.\label{dppost}
\end{equation}
From (\ref{Pncini}-\ref{dppost}) we obtain 
\begin{align}
P_{u}(t,t_{N})=\exp[-I(t)+I(t_{N})].\label{Pnc}
\end{align}

The probability $P_{u}$ gives also the fraction of points on the
wall of the bubble nucleated at $t_N$ which are still in the false vacuum at time $t$, 
i.e., the uncollided
fraction of its surface. The result is $P_{u}=P_{fv}(t)/P_{fv}(t_{N})$, which
has a simple interpretation. Consider a large volume $V$. Inside
this volume, a nucleation at time $t_{N}$ can only occur in the available
volume $VP_{fv}(t_{N})$. The nucleated bubble is initially
uncollided. For very large $V$, the probability that part of this single bubble 
leaves the volume $VP_{fv}(t_{N})$ at later times is negligible.
Nevertheless, this initial volume is invaded due to the nucleation and growth
of many other bubbles, and, by time $t$, a smaller part of it, $VP_{fv}(t)$,
remains in the false vacuum. Thus, the reference bubble is 
still contained in the initial volume but, in average, only a fraction 
$VP_{fv}(t)/VP_{fv}(t_{N})$ of its points
remains	 in the false vacuum region. This alternative derivation gives also 
the fraction of the bubble volume which is not covered by other bubbles.

\subsection{Probability that two points of a bubble wall remain uncollided}

We now consider two points $p$ and $p'$ on the surface of a bubble $B$
nucleated at time $t_{N}$. We shall first find the conditional probability
that $p$ remains in the false vacuum at time $t$, given that $p'$
was in the false vacuum at time $t'$. Following the same steps of 
Sec.~\ref{fv2}, we obtain again
\begin{equation}
P^S_{p|p'}=e^{-\int_{t_{c}}^{t}dP(t'')},\label{Pppini-1}
\end{equation}
(the superscript $S$ indicates that the two points belong to the surface of the bubble). 
We only need  to re-calculate the probability $dP(t'')$ of $p'$ being
outside of any bubble nucleated before $t''$ and inside a bubble
nucleated between $t''$ and $t''+dt''$. Such a bubble must have
nucleated within a sphere of radius $R(t'',t)$ centered at $p$
(the dotted region in previous figures). As we have already seen, under the above conditions
the dotted region is not affected by bubbles nucleated at times $t_{\mathrm{prev}}<t''$,
but we must exclude those nucleation points which would prevent the
nucleation of bubble $B$ at time $t_{N}$. Besides, since
we are also assuming that the point $p'$ is in the false vacuum at time
$t'$, we must also exclude nucleation points which would affect this event.

\begin{figure}[tbh]
\centering
\includegraphics[width=1\textwidth]{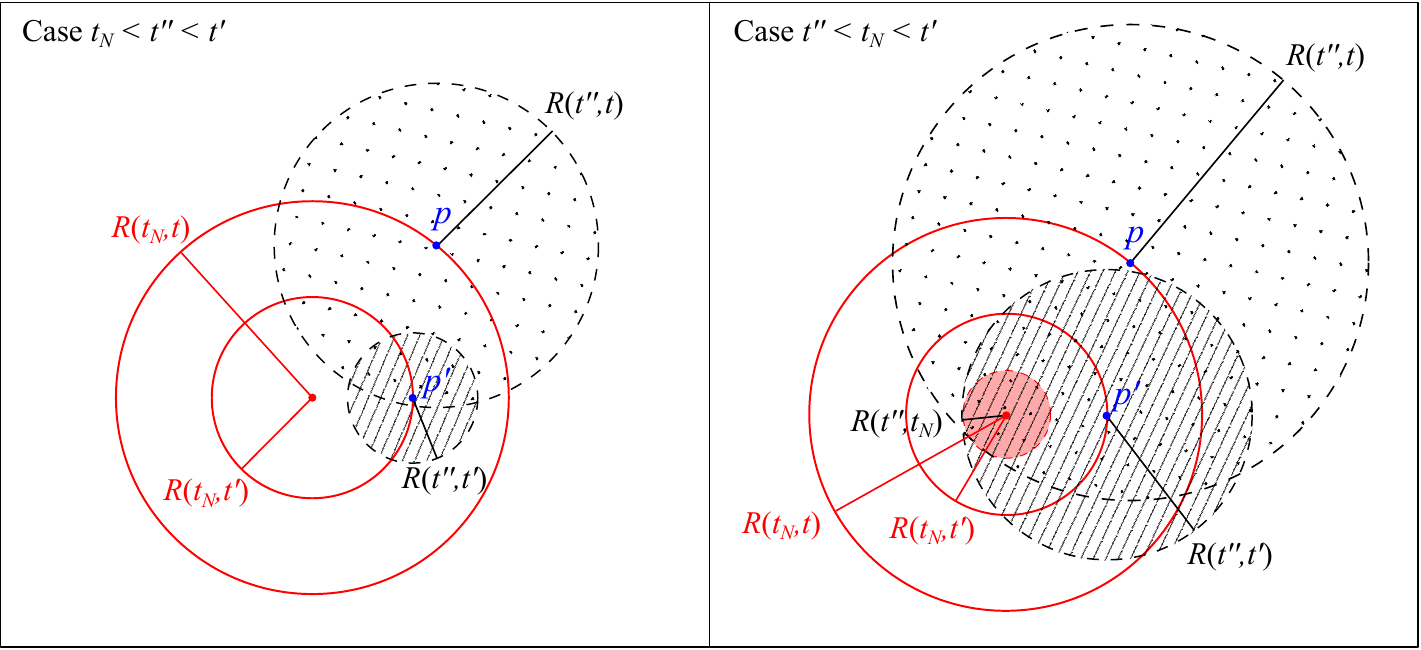}
\caption{The reference bubble $B$ nucleated at time $t_{N}$, at two subsequent
times $t'$ and $t$ (in red). The dots represent the nucleations
at time $t''$ which affect the point $p$ before time $t$. Nucleations
at $t''$ in the striped region would eat the point $p'$ before time
$t'$, and those in the shaded region would prevent the nucleation
of $B$.}
\label{figcorrel}
\end{figure}

For the sake of concreteness, let as assume that $t'\leq t$; the
case $t'>t$ is similar and gives essentially the same result\footnote{See footnote \ref{notatmaytp}.}.
For $t''>t'$, the nucleation at time $t''$ cannot affect events at
times $t_{N}$ or $t'$, so we have 
\begin{equation}
dP(t'')=dt''\Gamma(t'')\frac{4\pi}{3}R(t'',t)^{3}.\label{dppostt}
\end{equation}
The case $t''<t'$ is sketched in Fig.~\ref{figcorrel}. For $t''>t_{N}$
(left panel), a nucleation at time $t''$ cannot affect the nucleation of $B$
at time $t_{N}$ but may affect the point $p'$ before time $t'$. Hence,
any nucleation at time $t''$ must occur at a distance larger than
$R(t'',t')$ from $p'$ (i.e., outside the striped region). For $t''<t_{N}$
(right panel), a nucleation at $t''$ may also affect the nucleation
of the reference bubble. This will only happen if the nucleation at
$t''$ occurs within a radius $R(t'',t_{N})$ from the center
of $B$ (shaded region). Nevertheless, this region is fully contained in the
striped region\footnote{Since $R(t'',t_{N})+R(t_{N},t')=R(t'',t')$. In other words, 
the shaded region also affects the point $p'$ and is already taken into account.}.
Therefore, we only have to exclude the striped region from the dotted
one. Thus, for $t''<t'$ we have 
\begin{equation}
dP(t'')=dt''\Gamma(t'')\left[\frac{4\pi}{3}R(t'',t)^{3}-V_{\cap}(r,r',s)\right],\label{dpantt}
\end{equation}
where $V_{\cap}$ is the volume of the intersection of the dotted
and striped regions, given by Eq.~(\ref{VI}). 

From Eqs.~(\ref{Pppini-1}-\ref{dpantt}), we obtain the probability
of the point $p$ being in the false vacuum at time $t$ under the
condition that $p'$ is in the false vacuum at time $t'$,
\begin{equation}
P^S_{p|p'}(t,t',s)=\exp\left[-I(t)+I_{\cap}(t,t',s)\right],\label{Pcond}
\end{equation}
with $I_{\cap}$ given by Eq.~(\ref{Iint}). The result coincides
with Eq.~(\ref{Pcond1}), which corresponds to the case of two arbitrary
points in space. Here, the condition
that $p$ is attached to a bubble does not have more implications than the condition 
that $p'$ (on the same bubble) is uncollided\footnote{The
result would be different if $p'$ were a random point in space. Below we consider a similar case, namely, when $p'$ belongs to a different bubble $B'$.}.
Multiplying Eq.~(\ref{Pcond}) by the probability that $p'$ was uncollided at time $t'$, Eq.~(\ref{Pnc}), we obtain the joint
probability 
\begin{equation}
P^S_{p,p'}(t,t',t_{N},s)=\exp\left[-I(t)-I(t')+I(t_{N})+I_{\cap}(t,t',s)\right].\label{Pconj}
\end{equation}

\begin{figure}[tbh]
\centering
\includegraphics[height=5cm]{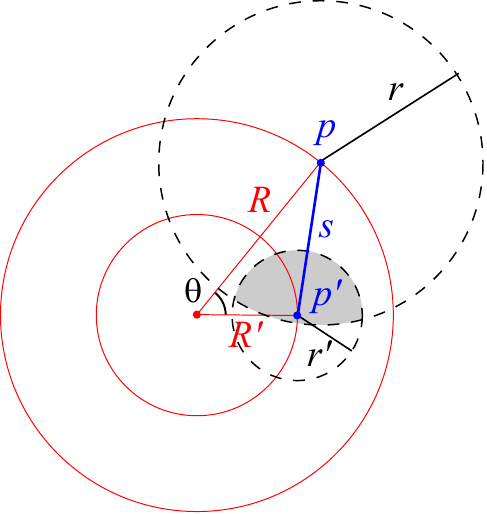}
\caption{The intersection volume $V_{\cap}$ and the separation $s$. The configuration
corresponds to the example on the left of Fig.~\ref{figcorrel}.}
\label{figcasq}
\end{figure}

As we have seen, the intersection volume $V_{\cap}$ depends on the distances $r=R(t'',t)$
and $r'=R(t'',t')$, and on the separation $s$. 
The latter can be written as a function of the bubble radii
\begin{equation}
R\equiv R(t_{N},t)\,,\quad R'\equiv R(t_{N},t'),\label{defradsR}
\end{equation}
and the angle $\theta$
between the positions of the points $p$ and $p'$ relative to the
bubble center (see Fig.~\ref{figcasq}), 
\begin{equation}
s=\sqrt{R^{2}+R^{\prime2}-2RR'\cos\theta}.\label{s}
\end{equation}
We thus have $R-R'\leq s\leq R+R'$. As we have seen in Sec.~\ref{fv2},
for $s\leq r-r'$ we have $V_{\cap}=4\pi r^{\prime3}/3$. In the present
case, in which the two points belong to the same bubble wall, we will
never actually have $s<r-r'$. Indeed, notice that 
\begin{equation}
r-r'=R(t',t)=R-R'\leq s.
\label{cotas}
\end{equation}
On the other hand, we may have $r+r'>s$, for which $V_{\cap}=0$,
so we write 
\begin{equation}
V_{\cap}=\frac{\pi}{12}(r+r'-s)^{2}\left[s+2(r+r')-\frac{3(r-r')^{2}}{s}\right]\Theta(r+r'-s).
\label{VIsb}
\end{equation}

\subsection{Points on walls of different bubbles}

Now we consider the case in which the two points $p$ and $p'$ belong
to the walls of two different bubbles $B$ and $B'$, nucleated at
times $t_{N}$ and $t_{N}^{\prime}$, respectively.

\subsubsection{General considerations}

There are some conditions which will have to be taken into account
eventually. In the first place, we assume that both reference bubbles
exist, so neither bubble should be occupying the nucleation center
of the other one. This implies that the distance $l$ between the
bubble centers must be larger than the distance travelled by a wall
from one center to the other, 
\begin{equation}
l>|R(t_{N},t_{N}^{\prime})|.\label{condsepar}
\end{equation}
In the second place, if the bubbles are too close, it may happen,
for instance, that the point $p'$ by time $t'$ is already inside the
bubble $B$. This case will be forbidden from the beginning when
we consider a conditional probability which assumes that $p'$ is  
uncollided at that time. On the other hand, when we consider
the joint probability for both points to be uncollided at
the corresponding times, the situation is not forbidden  
but its probability vanishes. We shall assume that we are not in this
situation, which implies the condition
\begin{align}
d'>R(t_{N},t'),\label{separdp}
\end{align}
where $d'$ is the distance from the point $p'$ to the center of the
bubble $B$. Similarly, requiring that the point $p$ is not inside
the bubble $B'$ by time $t$, we have the condition 
\begin{align}
d>R(t_{N}^{\prime},t),\label{separd}
\end{align}
where $d$ is the distance from $p$ to the center of $B'$. These
two conditions together imply Eq.~(\ref{condsepar})\footnote{Let us denote $\mathbf{l}$ the vector going from the center of $B$
to that of $B'$, $\mathbf{R}$ the vector joining the center of $B$
with $p$, and $\mathbf{R}'$ the vector joining the center of $B'$
with $p'$. We have $R=R(t_{N},t)$, $R'=R(t_{N}^{\prime},t')$, 
$d=|\mathbf{l}-\mathbf{R}|$, and $d'=|\mathbf{l}+\mathbf{R}'|$ (see Fig.~\ref{figcasq2}).
Then, the triangular inequality gives $d\leq l+R$ and $d'\leq l+R'$.
Inserting these inequalities in Eqs.~(\ref{separd}-\ref{separdp})
gives Eq.~(\ref{condsepar}).}. These restrictions do not affect the discussions 
on the nucleations
at time $t''$ below, and the examples shown in the figures 
fulfill them. Nevertheless, in applying our results, it should be
taken into account that the probability vanishes beyond the limits imposed by these conditions.

\subsubsection{Probability that a point of a bubble wall is uncollided, given
that a point of another bubble wall is uncollided}

\begin{figure}[tbh]
\centering
\includegraphics[width=1\textwidth]{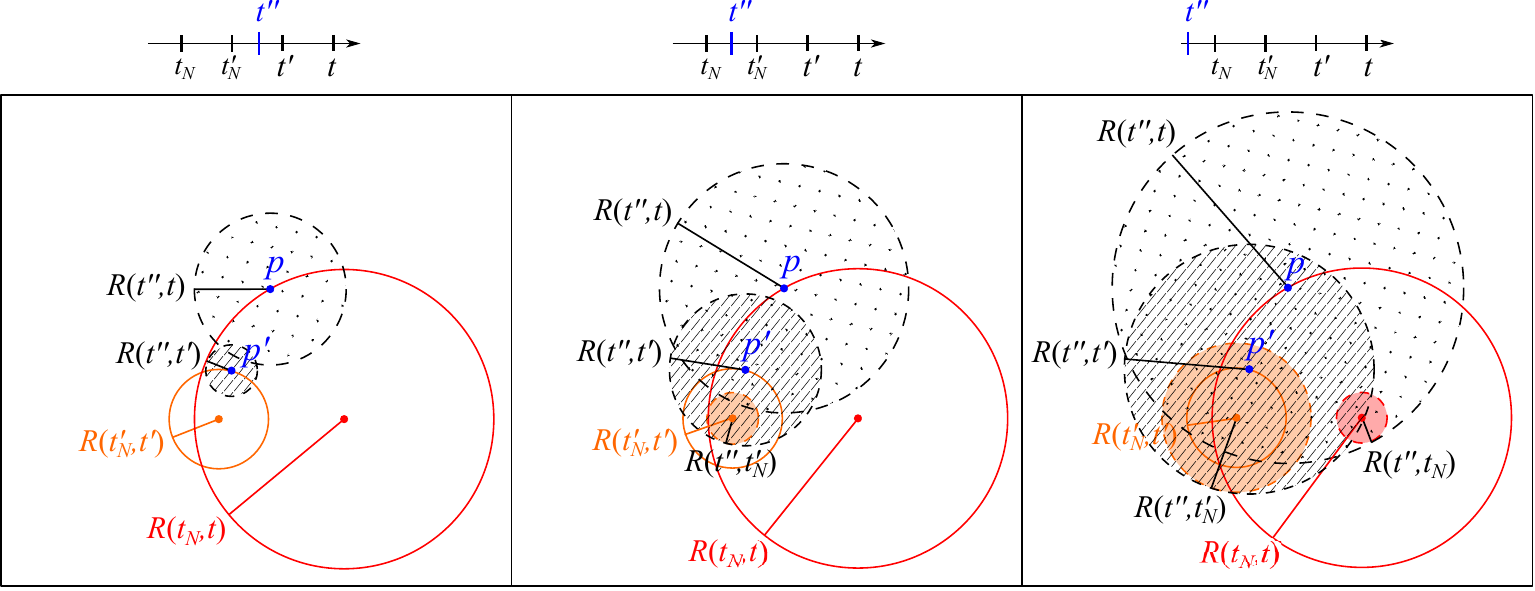}
\caption{A bubble B nucleated at time $t_{N}$ (red) and a bubble $B'$ nucleated
at time $t_{N}^{\prime}$ (orange). The former is drawn at time $t$
and the latter, at time $t'$, with the points $p$ and $p'$ on each
bubble surface. The black dots indicate the nucleations at time $t''$
which affect $p$ at time $t$. Those which fall inside the striped region would
also affect $p'$ at time $t'$, and those in the shaded regions would
affect the nucleations of $B$ or $B'$.}
\label{figcorr2b}
\end{figure}

First, we assume that $p'$ is uncollided at time $t'$, and we	
calculate the probability that $p$ is uncollided at
time $t$. For the sake of concreteness we shall consider only the 
case\footnote{See footnote \ref{notatmaytp}.}
$t\geq t'$, but we must  consider the two possibilities $t_{N}^{\prime}<t_{N}$
and $t_{N}<t_{N}^{\prime}$. Thus, there are three possible time orderings,
namely, $t_{N}^{\prime}<t'<t_{N}$, $t_{N}^{\prime}<t_{N}<t'$, or $t_{N}<t_{N}^{\prime}<t'$
(the latter is considered in Fig.~\ref{figcorr2b}). The conditional
probability is again given by 
\begin{equation}
P_{p|p'}^{SS'}(t,t')=e^{-\int_{t_{c}}^{t}dP(t'')}\label{Pppini-2}
\end{equation}
(the superscript $SS'$ indicates that the points belong to the 
surfaces $S$ and $S'$ of two different bubbles), and we must compute the probability $dP(t'')$ that $p$
has not been reached by bubbles nucleated before time $t''$ but
has been reached by a bubble nucleated in the interval $[t'',t''+dt'']$.
As before, the nucleation at $t''$ must occur within a sphere of
radius $R(t'',t)$ centered at $p$ (dotted region). However, some
of these nucleations will also affect the point $p'$ or the nucleations
of the bubbles $B$ or $B'$, and must be excluded.

Let us consider the time sequence $t_{N}<t_{N}^{\prime}<t'<t$. The other cases
are similar and lead to the same conclusion (see the appendix). Fig.~\ref{figcorr2b}
shows examples of the bubble configuration, corresponding to
particular positions of the time $t''$ relative to the other times
(shown in the timelines on top of each figure). The case $t''>t'$ is the simplest one and 
is not shown in Fig.~\ref{figcorr2b}.
In this case, the nucleation at $t''$ cannot
affect the events at times $t'$, $t_{N}^{\prime}$ or $t_{N}$, and
we have the whole dotted volume. Hence,
\begin{equation}
dP(t'')=dt''\Gamma(t'')\frac{4\pi}{3}R(t'',t)^{3}\qquad(t''>t').
\end{equation}

In the case $t_{N}^{\prime}<t''<t'$ (left panel of Fig.~\ref{figcorr2b}),
a bubble nucleated at time $t''$ may have eaten the point $p'$ by
time $t'$, so we must exclude the sphere of radius $R(t'',t')$ centered
at $p'$ (striped region). We thus have 
\begin{equation}
dP(t'')=dt''\Gamma(t'')\left[\frac{4\pi}{3}R(t'',t)^{3}-V_{\cap}\right]
\qquad(t_{N}^{\prime}<t''<t'),\label{tsegment}
\end{equation}
where $V_{\cap}$ is the volume of the intersection of the striped
and dotted spheres, which is given by Eq.~(\ref{VI}).

For $t_{N}<t''<t_{N}^{\prime}$ (shown in the central panel of Fig.~\ref{figcorr2b}),
the nucleation at $t''$ may \emph{also} prevent the nucleation of
bubble $B'$. Nevertheless, like in the previous section, the region
which can affect this event (light orange shade in Fig.~\ref{figcorr2b})
is completely contained within the striped region, which is already
excluded in Eq.~(\ref{tsegment}). Therefore, nothing changes when
$t''$ becomes smaller than $t_{N}^{\prime}$, 
\begin{equation}
dP(t'')=dt''\Gamma(t'')\left[\frac{4\pi}{3}R(t'',t)^{3}-V_{\cap}\right]
\qquad(t_{N}<t''<t_{N}^{\prime}).
\end{equation}

\begin{figure}[tbh]
\centering
\includegraphics[width=5cm]{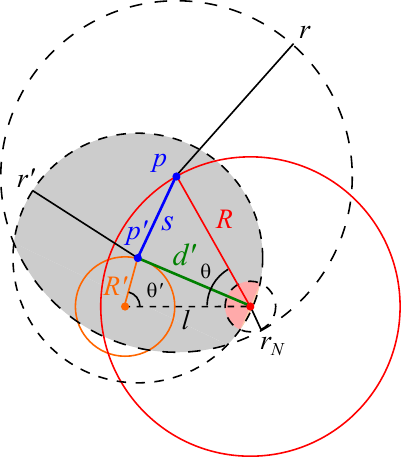}
\caption{The intersection volumes $V_{\cap}$ (gray) and $V_{\cap}^{\prime}$
(pink), and the distances $s$ and $d'$. This specific configuration
corresponds to the case on the right panel of Fig.~\ref{figcorr2b}.}
\label{figcasq2}
\end{figure}

Finally, for $t''<t_{N}$ (right panel), the nucleation at
$t''$ may \emph{also} prevent the nucleation of bubble $B$. Therefore,
we must exclude the sphere of radius $R(t'',t_{N})$ around
the center of $B$ (pink shade), as well as the striped region.
We thus have
\begin{equation}
dP(t'')=dt''\Gamma(t'')\left[\frac{4\pi}{3}R(t'',t)^{3}-\frac{4\pi}{3}R(t'',t_{N})^{3}-V_{\cap}+V_{\cap}^{\prime}\right]
\quad (t''<t_{N}).\label{tsegmentnpri}
\end{equation}
Here, we have first subtracted the volume of the pink region, which
is completely contained inside the dotted region, then we have subtracted
the intersection volume $V_{\cap}$ of the striped and dotted regions.
The volume  $V_{\cap}^{\prime}$ is a correction for the case in which 
the striped region overlaps with the pink region, like in the example of Fig.~\ref{figcorr2b}.
This volume must be added in order to avoid
subtracting twice their intersection.
This happens when the distance $d'$ between $p'$ and the center of $B$ is short enough. 
Thus, $V_{\cap}^{\prime}$  depends on $d'$ and on 
the radii of the two spheres,
$r'=R(t'',t')$ and $r_{N}\equiv R(t'',t_{N})$ (see Fig.~\ref{figcasq2}), and we 
have\footnote{The Heaviside function takes into account the fact that for large enough separation the intersection is empty. On the other hand, the condition (\ref{separdp}), $d'>R(t_N,t')=R(t'',t')-R(t'',t_N)=r'-r_N$, implies that the sphere of radius
$r_N$ will never be contained completely inside that of radius $r'$, except as a limit.}
\begin{equation}
V_{\cap}^{\prime}=\frac{\pi}{12}(r'+r_{N}-d')^{2}\left[d'+2(r'+r_{N})-\frac{3(r'-r_{N})^{2}}{d'}\right]
\Theta(r'+r_{N}-d').\label{Vintpri}
\end{equation}

Inserting these results in Eq.~(\ref{Pppini-2}), we obtain 
\begin{equation}
P_{p|p'}^{SS'}(t,t',s,d',t_{N}) =\exp\left[-I(t)+I(t_{N})+I_{\cap}(t,t',s)-I_{\cap}^{\prime}(t',t_{N},d')\right],
\label{Pcond2}
\end{equation}
where $I_{\cap}$ is the integral given by Eq.~(\ref{Iint}), and $I_{\cap}^{\prime}$ is a similar integral 
involving $V_\cap^\prime$. According to Eq.~(\ref{tsegmentnpri}), the upper limit of this integral is $t_N$. However, 
if we take into account the possibility $t'<t_{N}$
(not considered in the example used for this derivation), we must write (see the appendix for details)
\begin{equation}
I_{\cap}^{\prime}(t',t_{N},d')=\int_{t_{c}}^{\min\{t_{N},t'\}}dt''\Gamma(t'')V_{\cap}^{\prime}(r',r_{N},d').\label{Iintpri}
\end{equation}
The distance $s$ between $p$ and $p'$
is no longer given by Eq.~(\ref{s}). We may relate the  relevant
distances with the orientations of the points on each
bubble surface,
\begin{align}
s^{2} & = l^{2}+R^{2}+R^{\prime2}-2lR'\cos\theta'-2lR\cos\theta
  +2RR'(\sin\theta\sin\theta'\cos\phi-\cos\theta\cos\theta'), \label{s2} \\
d' & = \sqrt{R^{\prime2}+l^{2}-2R'l\cos\theta'}, 
\quad d= \sqrt{R^{2}+l^{2}-2Rl\cos\theta},\label{d}
\end{align}
where  $R\equiv R(t_{N},t)$, $R'\equiv R(t_{N}^{\prime},t')$,	
$l$ is the separation between the bubble centers,
the angles $\theta$ and $\theta'$ (which are in the interval $[0,\pi]$) correspond to the orientations
of the points $p$ and $p'$ on each bubble with respect to the axis
joining the two centers (see Fig.~\ref{figcasq2}), and $\phi$ (in the interval $[0,2\pi]$)
is the angle between the projections of these directions on the plane perpendicular
to the axis. Although Eq.~(\ref{Pcond2}) does not depend on $d$,
this distance appears in the condition (\ref{separd}).
Indeed, 
in the derivation of $P_{p|p'}^{SS'}$ we have assumed that the conditions (\ref{separdp}-\ref{separd}) are fulfilled. 
The assumption $d'>R(t_N,t')$ is correct, since the conditional probability assumes 
that the point $p'$ is uncollided. On the other hand, the condition $d>R(t_N^\prime,t)$ 
is not necessarily valid. If it is not fulfilled, the point $p$ at time $t$ is inside bubble $B'$, and
the probability just vanishes, so we must multiply Eq.~(\ref{Pcond2}) by the Heaviside function
\begin{equation}
\Theta\left(d-R(t_N^\prime,t)\right). \label{heaviside}
\end{equation}

It is interesting to consider the limit 
in which $B$ and $B'$ nucleate at the same time
and very close to each other. 
For $l= 0$, Eq.~(\ref{d}) gives $d'=R'$ 
and $s$ becomes the same as for the single-bubble case, i.e., 
Eq.~(\ref{s2}) becomes Eq.~(\ref{s}). 
Besides, for  $t_{N}=t_{N}^{\prime}$
we have $r'-d'=R(t'',t')-R(t_{N}^\prime,t')=R(t'',t_{N}^\prime)=r_N^\prime=r_N$.
Using this result, the volume $V_{\cap}^{\prime}$
becomes $V_{\cap}^{\prime}=\frac{4\pi}{3}r_N^{3}.$ Hence,
$I_\cap^\prime$ cancels with $I(t_{N})$  in Eq.~(\ref{Pcond2}),
and we obtain $P_{p|p'}^{SS'}(t,t',s)=\exp\left[-I(t)+I_{\cap}(t,t',s)\right]$,
which coincides with Eq.~(\ref{Pconj}),
i.e., the probability $P_{p|p'}^S(t,t',s)$ for two points on the same
bubble wall. This was to be expected, since in this limit the two bubbles are almost coincident. 
However, we must also take into account Eq.~(\ref{heaviside}). In particular, in this limit many points 
on each surface must be eaten by the other bubble (here, we are assuming that $p'$ is not).
In the case $t_N^\prime=t_N$  Eq.~(\ref{heaviside}) becomes 
$\Theta(d-R)$. For  $l\to 0$ we have $d\to R$, so we must be careful with the limit.
For $l\ll R$  Eqs.~(\ref{d})
can be written 
\begin{equation}
d-R=-l\cos\theta,\quad d'-R'=-l\cos\theta'. \label{daprox}
\end{equation}	
Hence, the Heaviside function \emph{vanishes} for $\cos\theta>0$, i.e., for $\theta<\pi/2$.
This is because,  in this limit, a half of bubble $B$
is inside $B'$.

\subsubsection{Probability that a point on a bubble wall is uncollided, in
the presence of another bubble}

To obtain the joint probability that a point $p$ on
the surface of $B$ and a point $p'$ on the surface of $B'$ remain
uncollided at times $t$ and $t'$, respectively, we only
have to multiply $P_{p|p'}^{SS'}$ by
the  probability 
that the point $p'$ on the wall of $B'$ is uncollided at time $t'$
(without any condition on the point $p$).
This probability was obtained in Sec.~\ref{fracsup} and is given by Eq.~(\ref{Pnc}).
However, the conditions are different in the present case, since we assume the 
existence of another bubble, $B$, at a certain distance from $B'$ (otherwise, we cannot ask whether the point $p$ on $B$ is uncollided). Therefore, we must consider the probability that $p'$ on $B'$ is uncollided at time $t'$, in the presence of the bubble $B$. This probability may be also of interest on its own.

Following the derivation of Sec.~\ref{fracsup},
we consider the region of nucleations at time $t''$ which affect
$p'$ at time $t'$ (in Fig.~\ref{figfracsup} this was the dotted region but 
in Fig.~\ref{figcorr2b} it is represented by a striped region). 
Like before, we need to exclude nucleations at $t''$  which affect the nucleation of the 
reference bubble $B'$ at $t_N^\prime$ (the orange region in Fig.~\ref{figcorr2b}), but
also those which prevent the nucleation of $B$ at $t_N$ (the pink region).
We thus obtain
\begin{equation}
P^u_{p'|B}(t',t_{N}^\prime,t_{N},d')=\exp[-I(t')+I(t_{N}^{\prime})+I_{\cap}^{\prime}(t',t_{N},d')].\label{Pnc2b}
\end{equation}
The first two terms in the exponent are like in  Eq.~(\ref{Pnc}). However, the probability that $p'$ (on the surface of $B'$) is uncollided depends also on its distance to the center of $B$ and the nucleation time of the latter. 
In this derivation we have assumed that Eq.~(\ref{separdp}) is fulfilled.
Therefore, 
Eq.~(\ref{Pnc2b}) does not take into account the possibility that bubble $B$ has eaten the point $p'$, 
and we must add the factor 
\begin{equation}
\Theta\left(d'-R(t_N,t')\right). \label{heavisidepri}
\end{equation}

\subsubsection{Probability that two points on the walls of different bubbles are
uncollided}

The joint probability that both points are uncollided is given
by the product of Eqs.~(\ref{Pnc2b}) and (\ref{Pcond2}),
\begin{equation}
P_{p,p'}^{SS'}(t,t',t_{N},t_{N}^{\prime},s)= \exp\left[-I(t)-I(t')+I(t_{N})+I(t_{N}^{\prime})+I_{\cap}(t,t',s)\right].\label{Pconj2}
\end{equation}
The integral $I_{\cap}^{\prime}(t',t_{N},d')$ has canceled
out, so this expression depends only on the point separation
and not on the bubble separation $l$. 
The result is very similar to the single-bubble probability,
Eq.~(\ref{Pconj}), except for the extra term $I(t_{N}^{\prime})$
in the exponent. 
However, we remark that 
if any of the conditions (\ref{separdp}-\ref{separd}) is not fulfilled, one of
the points has been eaten by the other bubble, 
and the probability actually vanishes. Therefore, Eq.~(\ref{Pconj2})
must be multiplied by the Heaviside functions 
\begin{equation}
\Theta(d-R(t_{N}^{\prime},t))\Theta(d'-R(t_{N},t')).\label{heavisides}
\end{equation}
In particular, these conditions imply that the point separation is grater than the distance 
traveled by a bubble wall 
between the times $t'$ and $t$, i.e., $s>R(t',t)$  (for $t'<t$), which
is essentially the same as Eq.~(\ref{cotas}), 
i.e.\ $r-r'<s$.\footnote{The 
condition (\ref{separd}) implies $d>R(t_N^\prime,t')+R(t',t)=R'+R(t',t)$. On the other hand, 
the triangular inequality gives $d<R'+s$. 
Comparing these two inequalities, we obtain $R(t',t)<s$. It is also worth noting that in this case 
we have $R(t',t)=r-r'\neq R-R'$.} 
Therefore, the volume $V_\cap$, which is generally given by Eq.~(\ref{VI}), can be written in the form (\ref{VIsb}).

To see the dependence with $l$, let us consider, for simplicity, 
the case $t_{N}^{\prime}=t_{N}$ and $t'=t$, so that
we have two bubbles of the same size. In this case, Eq.~(\ref{heavisides})
becomes $\Theta(d-R)\Theta(d'-R)$. For $l>R$, Eqs.~(\ref{d})
give $l-R<d<l+R$, and the same for $d'$. For $l>2R$, both  $d$ and $d'$ are larger than $R$,
so the Heaviside functions give a factor 1. Thus, for large $l$,
the probability is given by Eq.~(\ref{Pconj2}), which depends only
on the point separation $s$. On the other hand, for $l<2R$ the two
bubbles overlap, and some points will have zero probability of being uncollided
(depending on $d$ and $d'$).
For $l<R$, Eqs.~(\ref{d}) give $R-l<d<R+l$, and
for $l\ll R$ 
we have Eqs.~(\ref{daprox})
which, inserted in (\ref{heavisides}) imply
$\theta>\pi/2, \theta'>\pi/2$. This is because, as already discussed,
when the two bubbles almost coincide, 
a half of each bubble is inside the other bubble.

\section{Uncollided walls}

\label{somecases}

From the probabilities derived above we may obtain correlations between walls of different bubbles 
or different parts of a bubble wall.
For concrete computations, we shall use a couple of simple models for $\Gamma(t)$
and $v(t)$, which we motivate below. 

\subsection{Specific models}

In the first place, we shall assume that, after a bubble nucleates, 
its wall immediately reaches a terminal 
velocity\footnote{The time required to reach the terminal velocity is, in general, 
several orders of magnitude shorter than the total time of bubble expansion 
(see footnote \ref{notaTt}) and can be neglected. This holds also in the runaway case, 
where the terminal velocity is $v=1$. Even in the case of wall corrugation mentioned below, 
this is a good approximation if the wall deformations are treated as small perturbations.}.
Assuming a homogeneous temperature, the time dependence of $\Gamma(T)$ and of $v(T)$
is determined by the function $T(t)$. 
The simplest possible model consists of a constant $\Gamma$ and a constant $v$.
This model has been sometimes used as a simplification for 
computationally demanding approaches, such as those involving lattice calculations 
or other types of simulations (see, e.g., \cite{bkvv95}).
However, this model is hardly realistic, since the nucleation rate is very sensitive to the temperature,
and the latter always varies during the phase transition.

For detonations or runaway walls,
the reheating can be ignored until the end of the phase transition,
so $T(t)$ decreases according to the adiabatic expansion. In this case,
an exponentially growing nucleation rate is generally a good approximation.
Although the wall velocity increases do to the temperature decrease, it
can be assumed to be constant during the short time
of bubble growth. This gives the widely used model
\begin{equation}
\Gamma(t)=\Gamma_* e^{\beta t}, \quad R(t,t')=v(t'-t). \label{modelexp}
\end{equation}

For deflagrations, the evolution of $T$ is more
involved (even assuming a homogeneous reheating). In general, there
is a supercooling stage followed by a sudden reheating, 
after which a phase-equilibrium stage may occur. Due to the
high sensitivity of the nucleation rate, $\Gamma$
turns off as soon as the reheating begins. In this case, a reasonable
approximation is to assume that all bubbles nucleate at the moment $t_*$
at which $T$ reaches its minimum and  $\Gamma$
reaches its maximum \cite{ma05}. The wall velocity also decreases due to the
reheating, but its evolution is less simple. 
Nevertheless, depending on the parameters, the velocity variation may not be significant
and a constant velocity is a reasonable approximation (we shall consider other possibilities 
in the next section). 
In such a case, we have
\begin{equation}
\Gamma(t)=n_{b}\delta(t-t_{*}),\quad R(t,t')=v(t'-t).\label{model}
\end{equation}
This simple model has also been considered sometimes in time-demanding numerical computations
(see e.g. \cite{hhrw15}).

The two models described by Eqs.~(\ref{modelexp}) and (\ref{model}) give phase transitions of a very different nature.
For an exponentially growing nucleation rate, the growth of stable-phase domains is dominated by bubble nucleation,
while in the case of simultaneous nucleation the growth of the stable phase is dominated by bubble expansion. The 
kinematics of the former is thus characterized by the time scale $\beta^{-1}$, whereas for the latter the bubble density gives the characteristic length scale. It is therefore of interest to use these two cases as opposite examples. 
We shall now consider some basic quantities which are related to
the physical consequences of the phase transition, and in the next section we shall discuss some specific applications.

\subsection{The envelope of uncollided walls}

\label{Smedia}

For a given bubble of radius $R$, the locus
of its uncollided wall is a subset of the sphere of radius $R$. A
given point on the sphere is characterized by two angles $\theta,\phi$
by means of the parametrization $\mathbf{r}=R\hat{r}$, where 
$\hat{r}=(\sin\theta\cos\phi,\sin\theta\sin\phi,\cos\theta)$.
The uncollided wall can be characterized by the indicator or characteristic
function
\begin{equation}
1_{S}(\theta,\phi)=\begin{cases}
1 & \text{ if }\mathbf{r}\in S,\\
0 & \text{ if }\mathbf{r}\notin S.
\end{cases}
\end{equation}
Thus, the area of this bubble wall can be written in the 
form\footnote{We use the same notation $S$ for the locus of the uncollided
wall and its area.} 
\begin{equation}
S=R^{2}\int_{0}^{2\pi}d\phi\int_{0}^{\pi} d\theta\,\sin\theta\,1_{S}(\theta,\phi).\label{Su0}
\end{equation}
If we regard the characteristic function as a stochastic variable
and average over bubbles \emph{of the same radius}
$R$, we have 
\begin{equation}
\langle S\rangle =R^{2}\int_{0}^{2\pi}d\phi\int_{0}^{\pi}\sin\theta d\theta\,\left\langle 
1_{S}(\theta,\phi)\right\rangle .\label{Sncav}
\end{equation}
For each direction $\hat{r}$, we have two possible values of $1_{S}(\theta,\phi)$,
with probabilities  $P_{\hat{r}}(1)$ and $P_{\hat{r}}(0)=1-P_{\hat{r}}(1)$, and we have 
$\left\langle 1_{S}(\theta,\phi)\right\rangle =P_{\hat{r}}(1)$.
This  is the probability that the point represented
by $\hat{r}$ is uncollided, which is given by Eq.~(\ref{Pnc})
and is independent of the direction, $P_{\hat{r}}(1)=P_{u}(t,t_{N})=e^{-I(t)+I(t_{N})}$.
Thus,
the average uncollided area of a bubble of radius $R$ is given by
\begin{equation}
\langle S\rangle =4\pi R^{2}e^{-I(t)+I(t_{N})}.\label{Sumed}
\end{equation}
To obtain the total surface	 in a given volume $V$, we must
multiply Eq.~(\ref{Sumed}) by the number of bubbles of radius $R$ in this volume,
and then integrate over $R$. According to Eq.~(\ref{radio}),  the
bubbles of radius $R$ are those which were nucleated at the time $t_{N}(R,t)$ such
that $R=\int_{t_{N}}^{t}v(t'')dt''.$ Thus, at time $t$, the
bubbles which have radii between $R$ and $R+dR$ are those nucleated
between $t_{N}-dt_{N}$ and $t_{N}$. The number  of these
bubbles 
is
\begin{equation}
dN=\Gamma(t_{N})VP_{fv}(t_{N})dt_{N}.
\label{dN}
\end{equation}
Since $P_{fv}(t_{N})=e^{-I(t_{N})}$,
we have 
\begin{equation}
\langle S_{\mathrm{tot}}\rangle=\int dN\,\langle S\rangle =
V e^{-I(t)}\int_{t_{c}}^{t}dt_{N}\Gamma(t_{N})\,4\pi R(t_{N},t)^{2}.
\label{Stotav}
\end{equation}
As discussed in Sec.~\ref{intro}, for phenomena which depend on
the bubble walls, the important measure of progress (rather than the volume fraction $f_b$)
will be the fraction of uncollided wall area, $f_S(t)$, which is obtained by
dividing Eq.~(\ref{Stotav}) by
$V\int dN\,4\pi R^{2}$,
\begin{equation}
f_{S}(t)=\frac{e^{-I(t)}\int_{t_{c}}^{t}dt_{N}\Gamma(t_{N})R(t_{N},t)^{2}}{\int_{t_{c}}^{t}dt_{N}e^{-I(t_{N})}\Gamma(t_{N})R(t_{N},t)^{2}}.\label{fS}
\end{equation}

\begin{figure}[bth]
\centering
\includegraphics[width=0.49\textwidth]{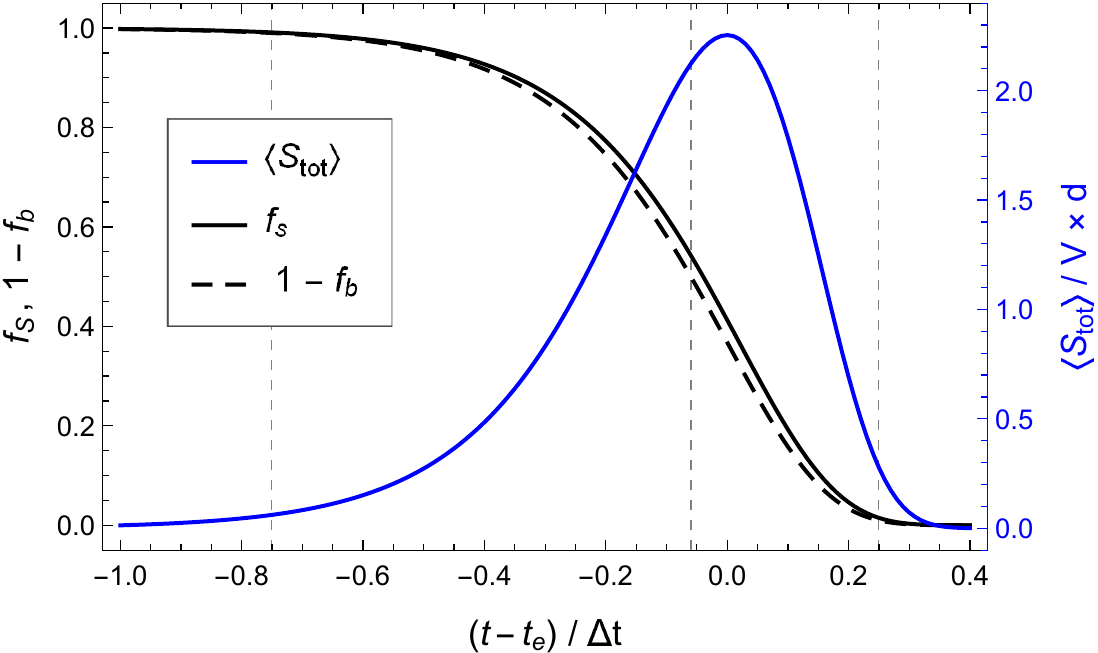}
\hfill
\includegraphics[width=0.49\textwidth]{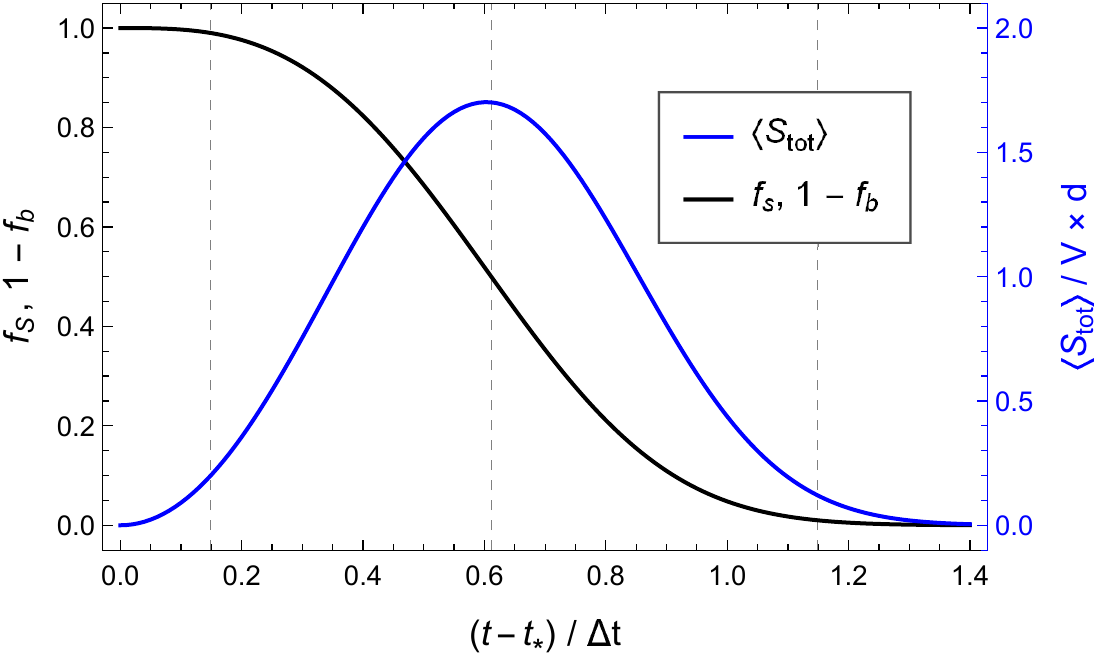}
\caption{The fraction of volume $P_{fv}=1-f_b$, the fraction of surface $f_S$, 
and the remaining wall area $\langle S_\mathrm{tot}\rangle$ at time $t$
for the exponential nucleation (left) and for the simultaneous nucleation (right). 
The vertical lines correspond to the values  $f_{b}=0.01$, $0.5$ and $0.99$.
\label{figsu}}
\end{figure}

In Ref.~\cite{tww92}, the energy-weighted fraction $f_{E}(t)$ is
also defined, by replacing $R(t_{N},t)^{2}$ with $R(t_{N},t)^{3}$
in the numerator and denominator of Eq.~(\ref{fS}).
It is to be expected that different measures of progress such as $f_{S}$,
$f_{E}$, or $P_{fv}=1-f_{b}$ (all of which vary from 1 to 0 throughout the phase transition) 
are qualitatively similar. In Ref.~\cite{tww92} it was found that, for
the case of an exponentially growing nucleation rate and a constant velocity, 
$P_{fv}(t)$ and $f_{E}(t)$ are very similar even quantitatively.
Let us consider the fraction of uncollided wall for this case. 
It is straightforward to calculate the integral (\ref{I}), which gives
$I(t)=8\pi v^3\beta^{-4}\Gamma(t)$. 
We define the time $t_e$  at which $P_{fv}$ has decreased to $1/e$, which
is given by the condition $I(t_e)=1$. 
Thus, we have $I(t)=e^{\beta (t-t_e)}$, and we obtain
\begin{equation}
\frac{\langle S_\mathrm{tot}\rangle}{V} = \frac{\beta}{v}e^{\beta (t-t_e)}
\exp\left[-e^{\beta (t-t_e)}\right].
\end{equation}
This quantity is proportional to the average nucleation rate $\bar{\Gamma}(t)=\Gamma(t)P_{fv}(t)$.
It is easy to see that the maximum of $\langle S_\mathrm{tot}\rangle$ 
(equivalently, the maximum of $\bar{\Gamma}$) 
occurs at $t=t_e$.
The parameter $\beta^{-1}$ characterizes the time scale for this transition, 
and, if we measure the length in units of the associated scale $v\beta^{-1}$, the quantity
$\langle S_\mathrm{tot}\rangle/{V}$ does not depend on the wall velocity.
In order to facilitate the comparison with the simultaneous nucleation model
considered below,
we shall also use sometimes a unit which
corresponds to a physical time in the evolution of the phase transition.  
Specifically, we are going to consider the interval $\Delta t$ between the moments at which 
$f_b=0.01$ and $f_b=0.99$. 
We have $\Delta t=c\beta^{-1}$, with $c=\log(\log0.01/\log0.99)\simeq 6.13$. 
Accordingly, we shall measure the area and the volume in units of the associated 
distance $d=v\Delta t$. 
In the left panel of Fig.~\ref{figsu} we plot $\langle S_{\mathrm{tot}}\rangle$,
$f_S$ and $P_{fv}$ as functions of time. 

For the delta-function rate (\ref{model}) we have $I(t)=n_{b}\frac{4\pi}{3}R(t_{*},t)^{3}$ and
\begin{equation}
\langle S_\mathrm{tot}\rangle=Vn_{b}\,4\pi R(t_{*},t)^{2}\, e^{-I(t)},
\label{avS0}
\end{equation}
while the denominator in Eq.~(\ref{fS}) is given by $Vn_{b}\,4\pi R(t_{*},t)^{2}$.
Therefore, we have $f_{S}(t)=e^{-I(t)}=P_{fv}(t)$. 
The same happens with $f_E(t)$; i.e., for simultaneous nucleation all these measures of progress coincide.
For this model, the natural unit of length is 
the characteristic distance $d_{b}\equiv n_{b}^{-1/3}$
(the ``average'' bubble separation),
and, for constant velocity,
the natural unit of time is the associated
value $t_{b}=d_{b}/v$ (which gives an estimate for the duration
of the phase transition). Thus, we may write 
\begin{equation}
f_{S}(t)=\exp\left[-\frac{4\pi}{3}\left(\frac{t-t_{*}}{t_{b}}\right)^{3}\right].
\end{equation}
In these units, this function does
not depend explicitly on the wall velocity.
The average wall area is given by 
\begin{equation}
\frac{\langle S_{\mathrm{tot}}\rangle}{V}=\frac{4\pi}{d_b} \left(\frac{t-t_{*}}{t_{b}}\right)^{2}\exp\left[-\frac{4\pi}{3}\left(\frac{t-t_{*}}{t_{b}}\right)^{3}\right].
\label{avS}
\end{equation}
To compare with the exponential nucleation model, we shall also use 
the units $\Delta t$ and $d=v\Delta t$
corresponding to the time interval between the values $f_b=0.01$ and $f_b=0.99$.   
Inverting the relation $I(t)=\frac{4\pi}{3}(R/d_b)^3$, we have 
$t-t_*=t_b[-\frac{3}{4\pi}\log(1-f_b)]^{1/3}$, and we obtain 
$\Delta t=\tilde{c} t_b$, with $\tilde{c}\simeq 0.9$.
We plot the functions $\langle S_\mathrm{tot}(t)\rangle$ and $f_S(t)$ in
the right panel of Fig.~\ref{figsu}.
	
We see that the width and the height of the curve of 
$\langle S_{\mathrm{tot}}\rangle$ is, in these units, quantitatively similar for the two models. 
The simultaneous case gives a somewhat lower and wider curve, 
but  such a precise comparison is not meaningful since these relations change 
if we use other units. For instance, if we use
the average bubble separation $d_b$ and the corresponding time scale $d_b/v$ for
the simultaneous case, we may define, 
for the exponential case, the bubble separation from the final bubble density, which gives
$d_b=(8\pi)^{1/3}\beta^{-1}$. It turns out that in these units the 
the curve becomes lower and wider for the exponential case.
This happens because the physical quantities $\Delta t$ or $d_b$,
which are appropriate for the model comparison, are defined from the 
dynamics of each phase transition and are not absolute units of time or length.

Some other common features are the following. 
When bubbles occupy a 1\% of space, the area in their walls is 
already relatively high. This is more visible in the simultaneous case, where 
$\langle S_{\mathrm{tot}}\rangle$ is more than a 10\% of its maximum value.
This is because of the high surface/volume ratio for small bubbles, which, moreover, are uncollided.
The maximum presence of walls occurs approximately in the
middle of the phase transition, when the fraction of volume
is $f_{b}\simeq0.63$ for the exponential case and $f_{b} \simeq0.49$ for the simultaneous case. 
Finally, when only a 1\% of space remains in the
false vacuum, the uncollided area is still relatively high.
This is more visible in the exponential case, where 
$\langle S_{\mathrm{tot}}\rangle$ is more than a 10\% of its maximum value. 
This relatively high value (compared to $f_S$) occurs because
$f_{S}$ is a fraction of an ever-increasing surface.
The main qualitative difference in these curves is that, 
in the simultaneous-nucleation case,
the evolution is quite symmetrical around the midpoint of the transition, 
while for the exponential rate, 
both the variation of $f_S$ and the maximum of $\langle S_{\mathrm{tot}}\rangle$ are
shifted towards the end of the transition.
This happens because the exponentially growing number of new (smaller) bubbles favors a higher total surface at later times. 
Nevertheless, 
in both cases, the maximum of $\langle S_{\mathrm{tot}}\rangle$ occurs 
near the time at which $f_b=0.5$

\begin{figure}[tbh]
\centering
\includegraphics[width=0.5\textwidth]{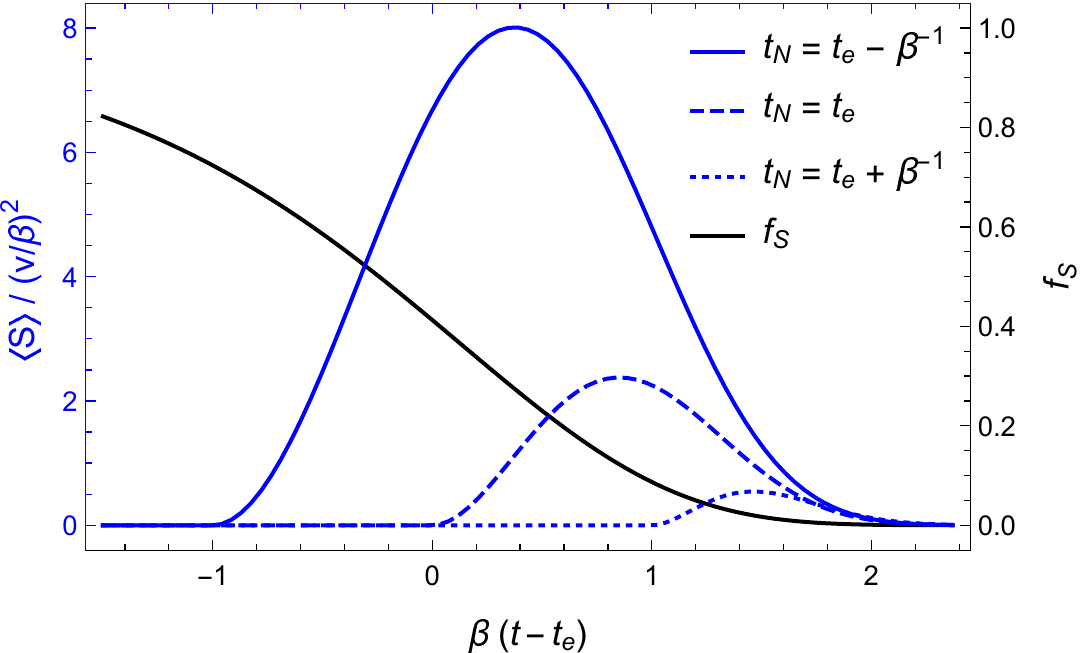} 
\hfill
\includegraphics[width=0.47\textwidth]{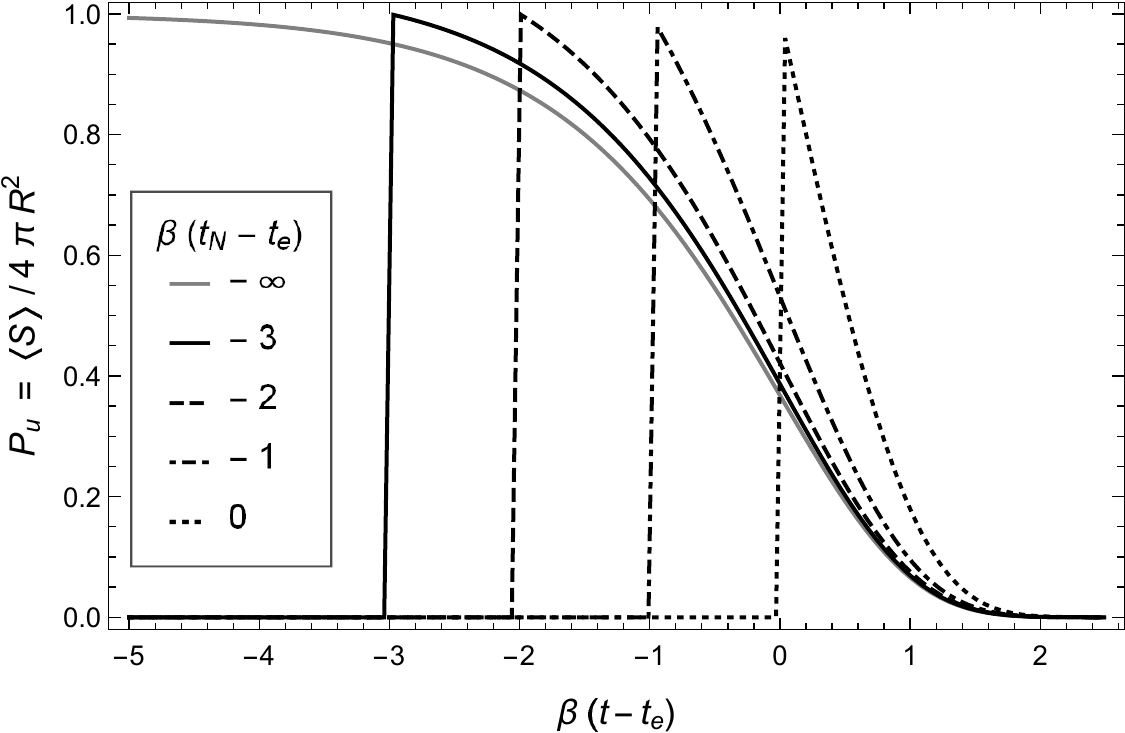}
\caption{Uncollided wall area of a bubble nucleated at time $t_N$ (left)
and fraction of uncollided wall (right) for the exponential case.
\label{figSexp}}
\end{figure}

For the delta-function rate, all the bubbles have the same size, and we have
$\langle S_{\mathrm{tot}}\rangle \propto \langle S\rangle$. 
In contrast, for the exponential rate, the size of a bubble depends on its nucleation time,
and the total average area is not proportional to the individual average area. 
In the left panel of Fig.~\ref{figSexp} we show the evolution of the individual average area
for this model for bubbles nucleated a different times
(we use the natural time unit $\beta^{-1}$, and we show the fraction of surface for reference). 
We see that the curves of $\langle S\rangle$ are more symmetrical than those of $\langle S_{\mathrm{tot}}\rangle$.
The average fraction of uncollided wall of the bubble, $\exp[-I(t)+I(t_{N})]$, 
coincides with
the probability that a single point on the wall remains uncollided, $P_{u}(t,t_{N})$.
For the delta-function nucleation rate, we have $t_{N}=t_{*}$ for all bubbles, and
we obtain $P_{u}=e^{-I(t)}=P_{fv}(t)$, which is already shown in the right panel of 
Fig.~\ref{figsu}. For the exponential rate, we show this quantity 
for a few values of $t_N$ in the right panel of Fig.~\ref{figSexp}.
The limit $t_N\to -\infty$ gives the volume fraction $P_{fv}(t)=1-f_b(t)$ (gray curve). 
As already mentioned,
the variation of this fraction occurs in a time $\Delta t\simeq 6.13\beta^{-1}$. 
For a bubble nucleated at $t_N\lesssim t_e-4\beta^{-1}$, its uncollided fraction is very close to this curve.
Bubbles which nucleate later are initially uncollided, but their uncollided wall fractions fall
faster, such that for $t=t_e+2\beta^{-1}$ they are all vanishingly small.

\begin{figure}[tbh]
\centering
\includegraphics[width=.49\textwidth]{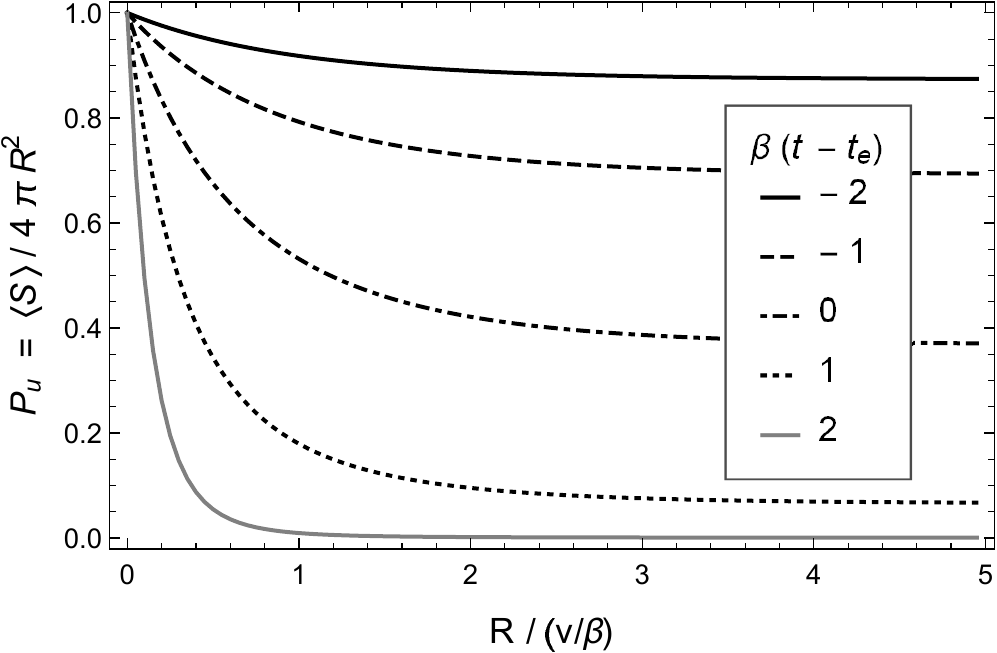} 
\hfill
\includegraphics[width=.49\textwidth]{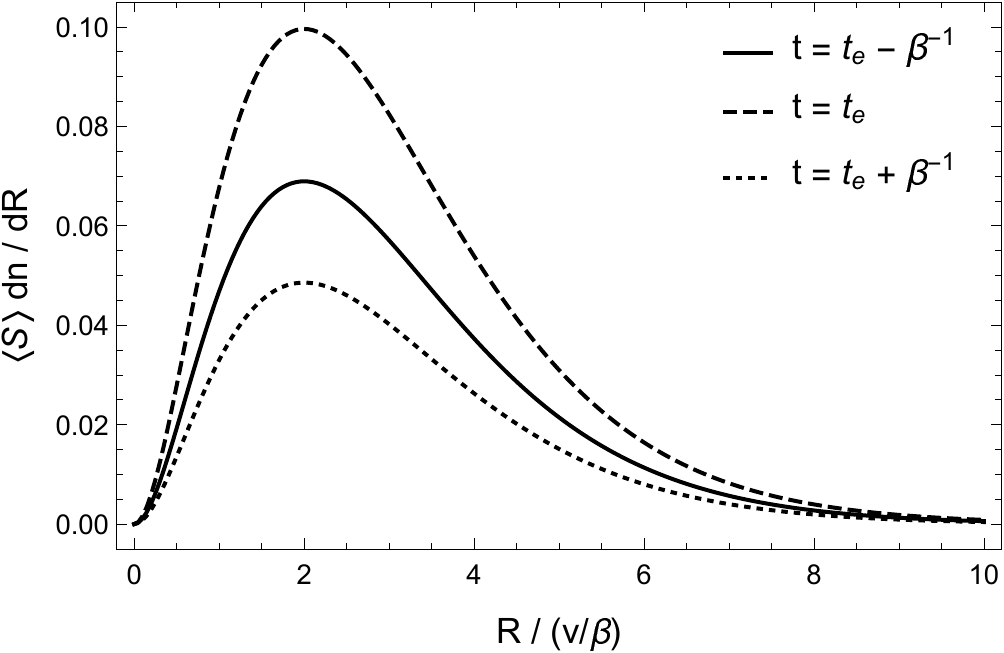}
\caption{Uncollided wall fraction of a bubble of radius $R$ (left) and
distribution of the total uncollided wall area in bubbles of a different sizes (right).
\label{figPuexp}}
\end{figure}

At a given time  $t$, $P_{u}(t,t_{N})$ gives the uncollided wall fraction of a bubble of 
radius  $R=v(t-t_N)$.
In the left panel of  Fig.~\ref{figPuexp} we show this fraction as a function of $R$ 
at different times $t$.
Very small bubbles (which nucleated very recently) are completely uncollided, 
while larger bubbles have a fraction of their wall already collided.
For a very large bubble (i.e., for $t_N\to -\infty$), 
its uncollided wall fraction is just given by $P_{fv}(t)$.
We see that, at early times (black solid line), bubbles of all sizes are almost uncollided, 
while at later times (gray solid line) the walls of all but the smallest bubbles 
have completely disappeared.

In this model there is a wide range of bubble sizes, and
a quantity of interest is the size distribution $dn/dR=V^{-1}dN/dR$. 
It is easily obtained using $dN$ from Eq.~(\ref{dN})
and $|dR|=v(t_{N})|dt_{N}|$, where, for constant velocity,  the nucleation time of a bubble of size $R$ is
given by $t_N=t-R/v$. We thus have
${dn}/{dR}= {\bar{\Gamma}(t-R/v)}/{v}$,
where $\bar{\Gamma}$ is the average nucleation rate ${\Gamma(t_{N})P_{fv}(t_{N})}$. 
Hence, the size distribution has always the same shape and is only shifted to higher $R$ at later times.
For the exponential nucleation rate, the maximum is at 
$R_{\max}=v(t-t_e)$ and takes the constant value
$dn/dR|_{\max}=(\beta/v)^4/8\pi e$.
Another quantity of interest is the
volume-weighted distribution of bubble sizes, $R^{3}dn/dR$,
which is associated to the energy that has
been released in bubbles of a given size. 
These quantities have been already discussed in the literature (see, e.g., \cite{tww92}).
The surface-weighted size distribution $R^{2}dn/dR$ may also be of
interest, depending on the application. In this case, it may be more appropriate to 
use the {uncollided}
surface as a weight, $\langle S\rangle{dn}/{dR}$, which for this model is given by
\begin{equation}
\langle S\rangle\frac{dn}{dR}
=\frac{1}{2} \frac{\beta^4}{v^4} I(t)e^{-I(t)} R^2 e^{-\beta R/v}.
\end{equation}
This distribution is shown at three times around $t_e$ in the right panel of 
Fig.~\ref{figSexp}. We see that it has always the same shape and only changes its amplitude, which is proportional to $\langle S_\mathrm{tot}(t)\rangle$.
The maximum is at the fixed value $R=2v/\beta$. 
Thus, the normalized distribution 
\begin{equation}
\frac{1}{\langle S_\mathrm{tot}\rangle}\frac{d\langle S_\mathrm{tot}\rangle}{dR}=
\frac{\langle S\rangle}{\langle S_\mathrm{tot}\rangle}\frac{dN}{dR}
=\frac{1}{2} \frac{\beta^3}{v^3} R^2 e^{-\beta R/v} 
\end{equation}
does not depend on time.

\subsection{Time correlations in the envelope}

The function $\langle S_{\mathrm{tot}}(t)\rangle$ describes the turning on and off 
of the system of walls which sources several of the consequences of the phase transition. 
However, as already mentioned, in some cases
the relevant quantity will be the time correlation
$\langle S_{\mathrm{tot}}(t)S_{\mathrm{tot}}(t')\rangle$, or even correlations between individual bubbles or
between parts of bubbles.
The total surface involves a sum over
individual bubbles, $S_{\mathrm{tot}}=\sum_{i}S_{i}$, and we may write
\begin{equation}
S_{\mathrm{tot}}(t)S_{\mathrm{tot}}(t')=\sum_{i}S_{i}(t)S_{i}(t')
+\sum_{i}\sum_{j\neq i}S_{i}(t)S_{j}(t').
\label{SSptot}
\end{equation}
Taking the ensemble average, the terms in the first sum involve time correlations
of a single bubble. These terms will depend only on the times
$t$, $t'$ and on the nucleation time $t_{N}$, but not on the bubble position. Therefore,
in a volume $V$, we may evaluate the sum by replacing $\sum_{i}\to V\int\Gamma(t_{N})P_{fv}(t_{N})dt_{N}$,
like we did in Sec.~\ref{Smedia}. 
On the other hand, the terms in the double sum
in Eq.~(\ref{SSptot}) will depend on the bubble separation $l$. Therefore, the sum
over $j$ can be replaced by the integral $4\pi\int dl\,l^{2}$.
The result of this integral does not depend on the bubble positions.
Then, the sum over $i$ can be evaluated as before. 
We shall study these two contributions separately.

Let us first consider a given bubble at two different times $t,t'$. 
The uncollided area at each time is given by Eq.~(\ref{Su0}), and we have
\begin{equation}
\langle S(t)S(t')\rangle =R^{\prime2}R^{2}\int_{0}^{2\pi}d\phi'\int_{0}^{\pi}\sin\theta'd\theta'
\int_{0}^{2\pi}d\phi\int_{0}^{\pi}\sin\theta d\theta\,\langle1_{S(t)}(\theta,\phi)1_{S(t')}(\theta',\phi')\rangle.
\label{SSp}
\end{equation}
The angles correspond to directions $\hat{r},\hat{r}'$ indicating points $p,p'$ on the surfaces $S(t)$ and $S(t')$, respectively.
For each pair $\hat{r},\hat{r}'$, the product
$1_{S(t)}1_{S(t')}$ takes the value $0$ or $1$, the
latter with probability $P_{\hat{r},\hat{r}'}(1)=P^S_{p,p'}(t,t',t_{N},s)$ given by Eq.~(\ref{Pconj}).
This probability depends on a single angle, namely, 
that between $\hat{r}$ and $\hat{r}'$. 
Using the relation (\ref{s}) for the point separation $s$ as a function of the angle between
 $\hat{r}$ and  $\hat{r}'$, 
we obtain
\begin{equation}
\langle S(t)S(t')\rangle=8\pi^{2}RR^{\prime}e^{-I(t)-I(t')+I(t_{N})}\int_{R'-R}^{R'+R}ds\,s\,e^{I_{\cap}(t,t',s)},\label{SSpriav}
\end{equation}
where $I_{\cap}$ is given by Eq.~(\ref{Iint}). 
If we consider two different bubbles, whose centers are a distance
$l$ apart, we may repeat the same steps which lead to Eq.~(\ref{SSpriav}). 
The only difference is that now we have two 
surfaces belonging to different bubbles, so we replace $S(t')$ with $S^{\prime}(t')$ in 
Eq.~(\ref{SSp}), and the 
probability $P_{\hat{r},\hat{r}'}(1)$ is given by Eqs.~(\ref{Pconj2}-\ref{heavisides}). We thus have
\begin{equation}
\langle1_{S(t)}(\theta,\phi)1_{S'(t')}(\theta',\phi')\rangle=P_{p,p'}^{SS'}(t,t',t_{N},t_{N}^{\prime},s)
\Theta(d-R(t_{N}^{\prime},t))\Theta(d'-R(t_{N},t')).
\end{equation}
We obtain 
\begin{align}
\langle S(t)S^{\prime}(t')\rangle= & \,2\pi R^{2}R^{\prime2}
e^{-I(t)-I(t')+I(t_{N})+I(t_{N}^{\prime})}
\int_{0}^{2\pi}d\phi\int_{0}^{\pi}d\theta\sin\theta\int_{0}^{\pi}d\theta'\sin\theta'\nonumber
 \\
 & \times 
 \exp[I_{\cap}(t,t',s)]\,\Theta(d-R(t_{N}^{\prime},t))\Theta(d'-R(t_{N},t')).
 \label{SSp2}
\end{align}
We could also change the variables of integration from the angles to the distances 
$s,d,d'$ through Eqs.~(\ref{s2}-\ref{d}). The result depends on the bubble separation $l$.

\subsubsection{Simultaneous nucleation}

For a delta-function rate we have $t_{N}=t_{*}$,
$I(t_{N})=0$, and $I_{\cap}=n_{b}V_{\cap}$, with 
\begin{equation}
V_{\cap}=\frac{\pi}{12}(R+R'-s)^{2}\left[s+2(R+R')-\frac{3(R'-R)^{2}}{s}\right]. \label{Vintsimult}
\end{equation}
In the two-bubble case, 
we have also $t'_{N}=t_{*}$ and $I(t'_{N})=0$. Besides, the Heaviside functions
become $\Theta(d-R)\,\Theta(d'-R')$. 
At equal times we have
$R'=R$, and the expressions become simpler. In particular, we have
\begin{equation}
V_{\cap}=\frac{\pi}{12}(2R-s)^{2}\left(s+4R\right).\label{Vintsimulteqt}
\end{equation}
In any case, the integrals in Eqs.~(\ref{SSpriav}) and (\ref{SSp2}) must be computed numerically.

\begin{figure}[tbh]
\centering
\includegraphics[height=8cm]{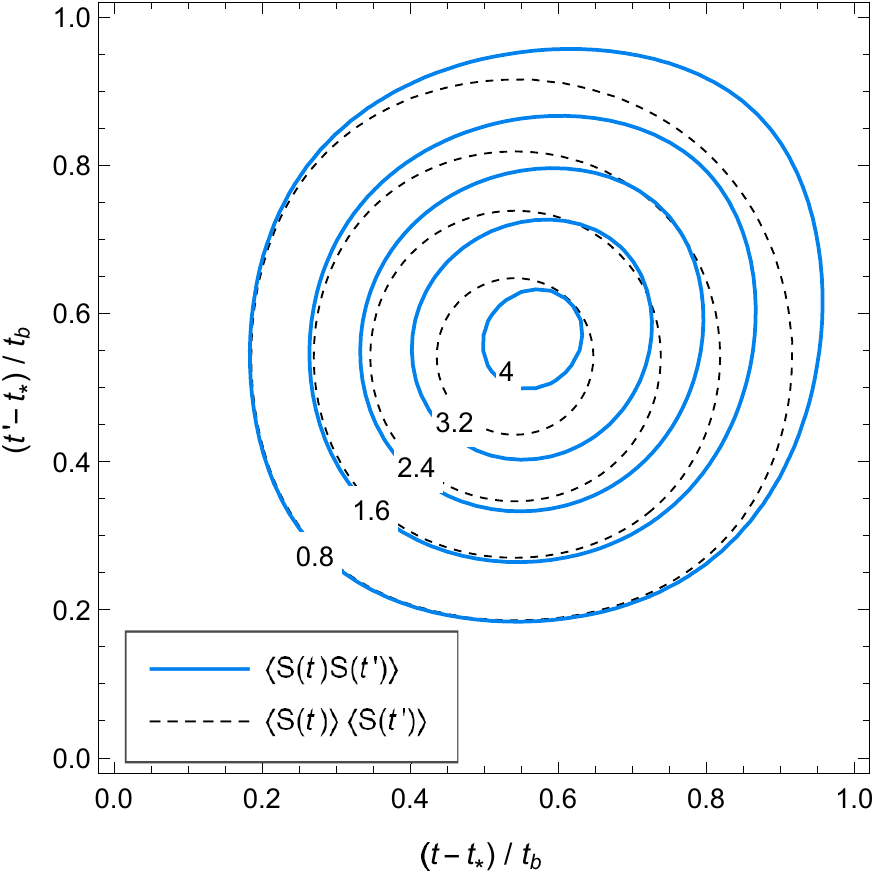} 
\hfill
\includegraphics[height=8cm]{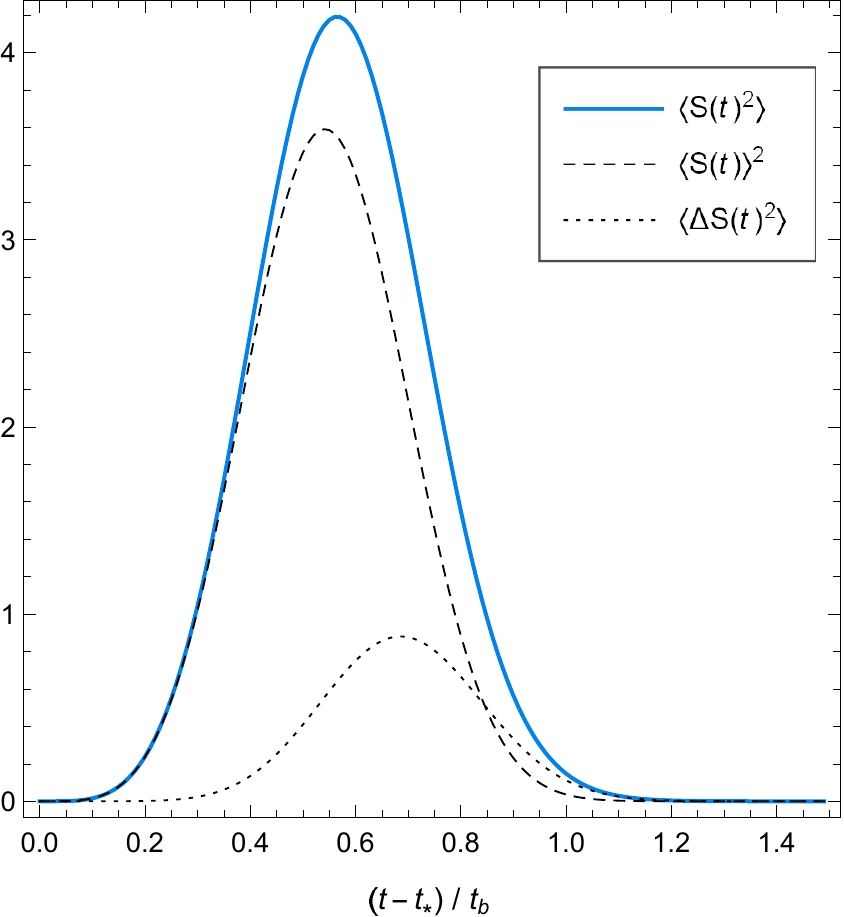}
\caption{Surface correlation of a bubble between two times for a delta-function rate, in units of $d_{b}^{4}$. 
Left panel: contours of $\langle S(t)S(t')\rangle$ and $\langle S(t)\rangle\langle S(t')\rangle$. 
Right panel: the equal-time values and their
difference. \label{figrms}}
\end{figure}

\begin{figure}[tbh]
\centering
\includegraphics[height=7.7cm]{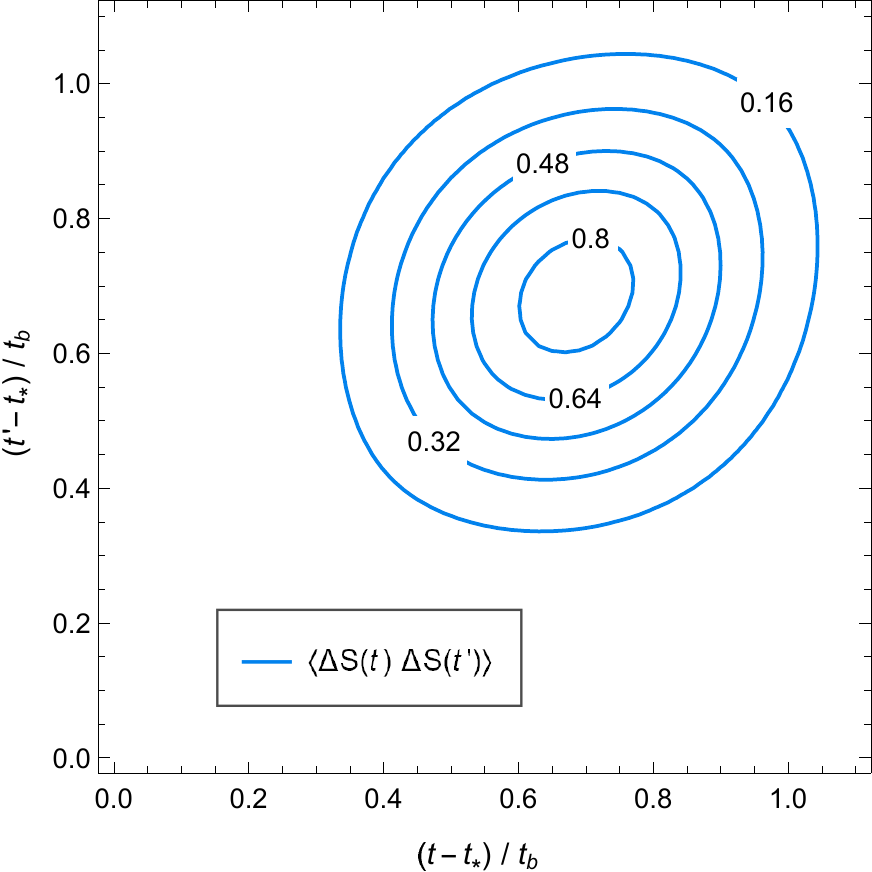}
\hfill
\includegraphics[height=7.7cm]{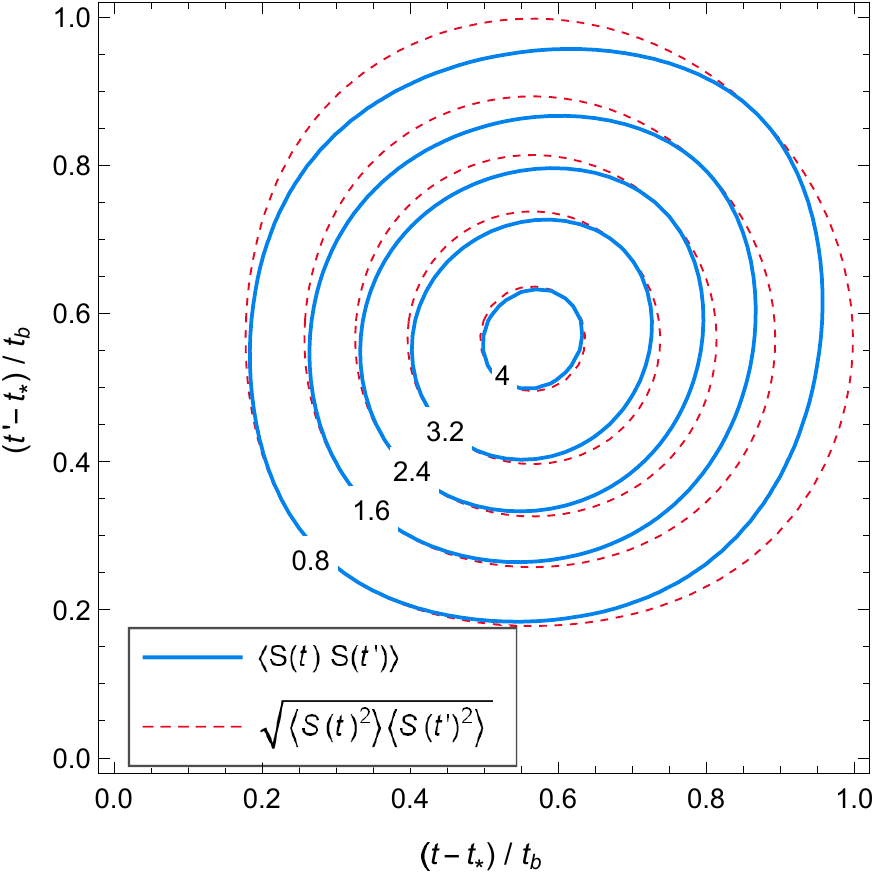} 
\caption{Surface correlation of a bubble between different times for a delta function rate.
Left panel: contours of $\langle\Delta S(t)\Delta S(t')\rangle$. 
Right panel: contours of $\langle S(t)S(t')\rangle$ and $\sqrt{\langle S(t)^{2}\rangle\langle S(t')^{2}\rangle}$.
The areas are in units of $d_b^2$.
\label{figcontour}}
\end{figure}

In Figs.~\ref{figrms} and \ref{figcontour} we consider the 
surface correlations for a single bubble.
In Fig.~\ref{figrms} we compare the functions $\langle S(t)S(t')\rangle$
and $\langle S(t)\rangle\langle S(t')\rangle$. 
The latter
has a quite simpler expression, since we have (in this model) 
$\langle S\rangle=4\pi R^2 e^{-I}$,
and could be used as an approximation for the former. Such an approximation
corresponds to assuming that the two surfaces are uncorrelated. We
see that these quantities are quite similar. In particular, the approximation is very good initially
(i.e., for small values of $t-t_{*}$ and $t'-t_{*}$). However, they
depart at later times, where the uncorrelated function
tends to zero more rapidly. In the right panel, the variance 
$\langle\Delta S^{2}\rangle$, where $\Delta S=S-\langle S\rangle$,
is also shown. 
The left panel of Fig.~\ref{figcontour} 
shows the covariance $\langle\Delta S(t)\Delta S(t')\rangle$.
In the right panel
we compare the function
$\langle S(t)S(t')\rangle$ with  $\sqrt{\langle S(t)^{2}\rangle\langle S(t')^{2}\rangle}$.
The latter is the result we would obtain if $S(t)$
and $S(t')$ were maximally correlated. Since Eqs.~(\ref{SSpriav}-\ref{Vintsimult})
are simpler for equal times, this function can also be used as
an approximation for the former. By definition, both coincide at $t=t'$,
so this approximation is better than the uncorrelated one 
at later times. On the other hand, it deviates for large $|t-t'|$.

\begin{figure}[tbh]
\centering
\includegraphics[width=0.49\textwidth]{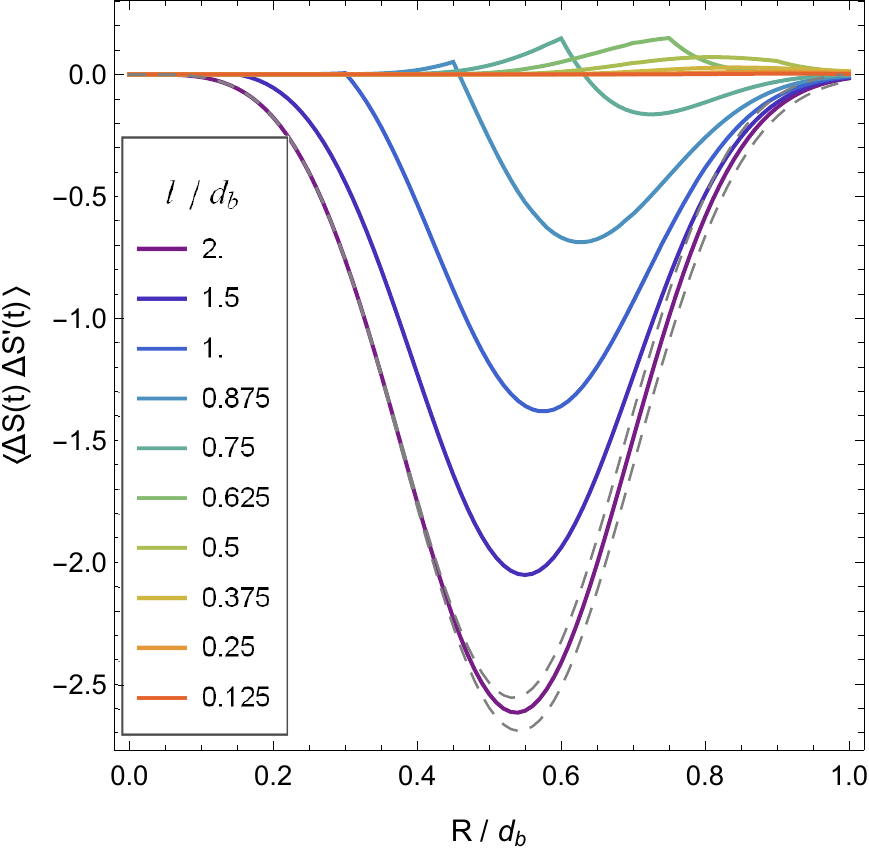} 
\hfill
\includegraphics[width=0.49\textwidth]{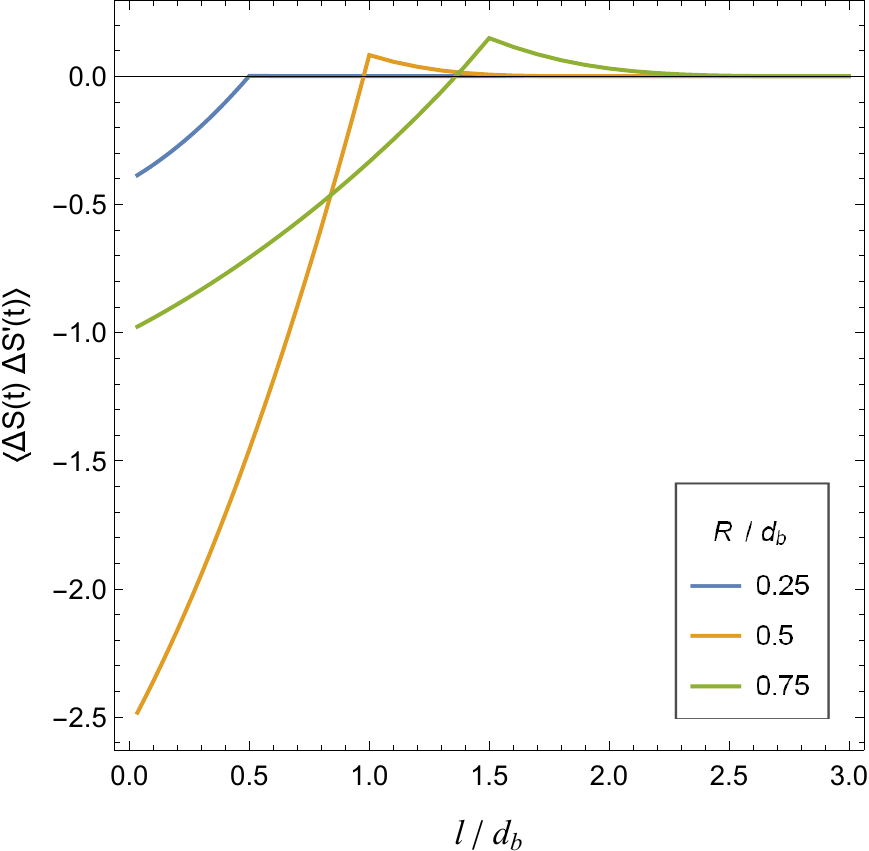}
\caption{The equal-time function $\langle\Delta S(t)\Delta S^{\prime}(t)\rangle$ for a delta-function rate
(in units of $d_{b}^{4}$) for two bubbles separated a distance $l$,
as a function of the bubble radius (left panel) and as a function
of the bubble separation (right panel). The dashed curves in the left
panel correspond to the functions $\langle S(t)^{2}\rangle/4-\langle S(t)\rangle^{2}$
(upper curve) and $-3\langle S(t)\rangle^{2}/4$ (lower curve).
\label{figneq}}
\end{figure}

In  Fig.~\ref{figneq} we show the result for two bubbles separated a distance $l$. 
We consider the deviations
$\Delta S, \Delta S'$, and we plot only the equal-time case.
In the left panel, the covariance\footnote{In the general case we have 
$\langle\Delta S(t)\Delta S^{\prime}(t')\rangle = \langle S(t)S^{\prime}(t')\rangle
-\langle S(t)\rangle\langle S^{\prime}(t')\rangle$. 
Notice that $\langle S^{\prime}(t')\rangle = \langle S(t')\rangle$. 
Thus, for equal times we have 
$\langle\Delta S(t)\Delta S^{\prime}(t)\rangle = \langle S(t)S^{\prime}(t)\rangle
-\langle S(t)\rangle ^2$.}  
$\langle\Delta S(t)\Delta S^{\prime}(t)\rangle$ 
is plotted as a function of the bubble
radius $R$, or, equivalently, as a function of time, since we have 
$R/d_{b}=(t-t_{*})/ t_{b}$.
The curves of different colors correspond to various values of the bubble
separation $l$. We see that the correlation vanishes for large $l$,
i.e., we have $\langle S(t)S^{\prime}(t)\rangle\to\langle S(t)\rangle\langle S(t)\rangle$
for $l\to\infty$. On the other hand,
the maximal correlation is attained for $l\to0$. This is also
appreciated in the right panel, which shows the covariance as a function
of $l$ for a few values of $R$. In all these curves, there is a
sudden change in the behavior at the point $l=2R$, i.e., when the
two bubbles come into contact. For $l<2R$, the two bubbles overlap,
so a part of their surface is surely collided. One could expect
that for $l\to0$ the quantity $\langle S(t)S^{\prime}(t)\rangle$
would match the value $\langle S(t)S(t)\rangle$. However,
in this limit, half of each bubble is surely collided, so a better guess would
be $\langle S(t)^{2}\rangle/4$. This value is indicated by the
upper dashed line in the left panel of Fig.~\ref{figneq}. We see
that this curve does not coincide with the
limit $l=0$ for different bubbles. The lower dashed line corresponds
to the approximation $\langle S(t)\rangle^{2}/4$.

\subsubsection{Exponential nucleation}

\begin{figure}[tbh]
\centering
\includegraphics[height=8cm]{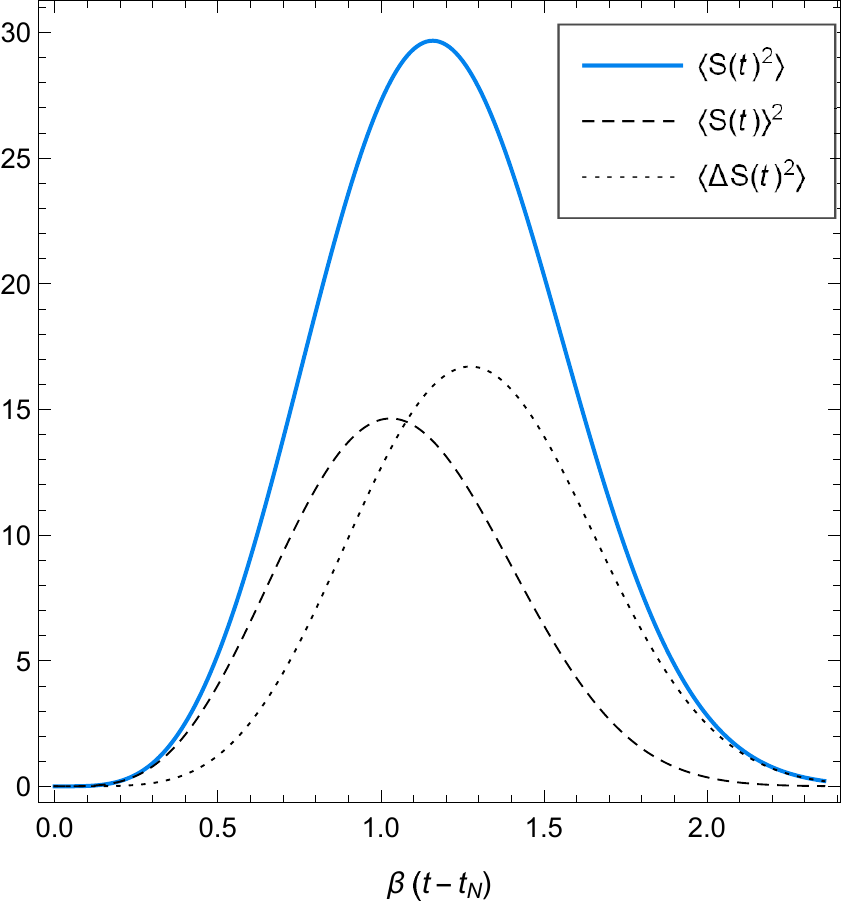} 
\hfill
\includegraphics[height=8cm]{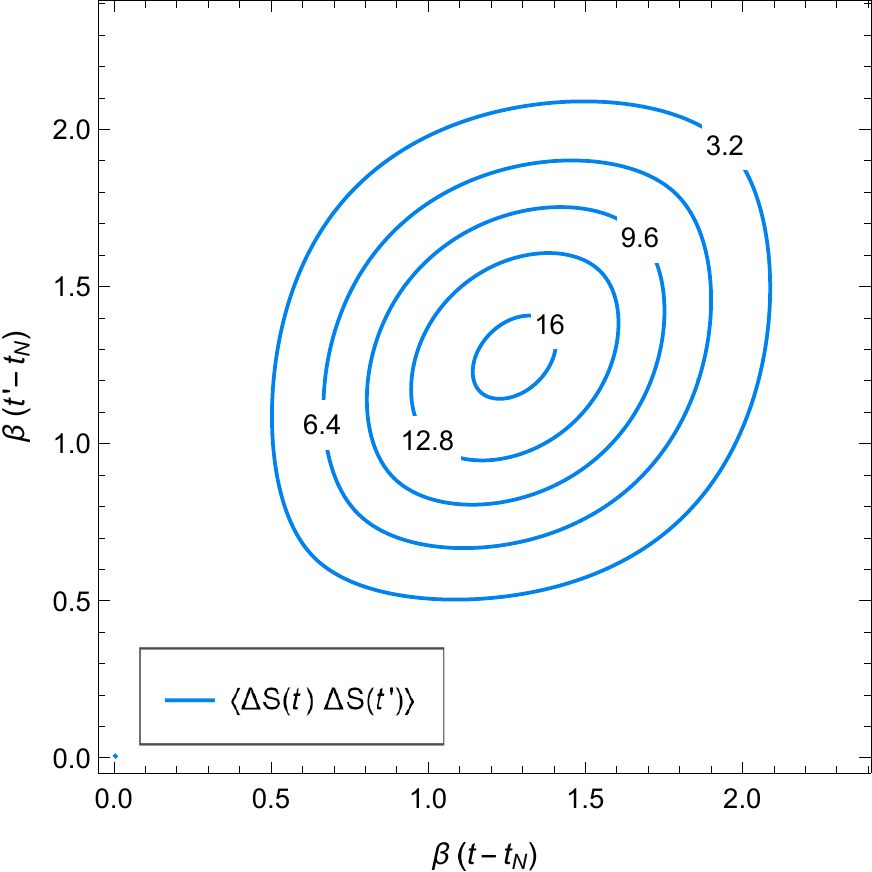}
\caption{Surface correlation  in the case of an exponential nucleation rate, 
for a bubble nucleated at time $t_N=t_\mathrm{med}$ such that $f_b(t_\mathrm{med})=0.5$. 
The area is in units of $(v/\beta)^{2}$. 
Left panel: equal-time values of $\langle S(t)S(t')\rangle$, $\langle S(t)\rangle\langle S(t')\rangle$, 
and $\langle\Delta S(t)\Delta S(t')\rangle$. 
Right panel:  contours of $\langle\Delta S(t)\Delta S(t')\rangle$.
\label{figrmsexp}}
\end{figure}

For the exponential nucleation rate, the general characteristics are qualitatively similar to the delta-function rate, 
as already seen for $\langle S_\mathrm{tot}\rangle$. 
Let us consider, for instance,
the wall area correlation for an {individual} bubble, which is given by Eq.~(\ref{SSpriav}).
For the integral (\ref{Iint})
we obtain\footnote{We must use the expression  (\ref{VIsb}) for 
the function $V_{\cap}$, where 
only  $r+r'=v(t+t'-2t'')$ depends on the integration variable $t''$.
The Heaviside function $\Theta(r+r'-s)$  gives the condition $t''<(t+t')/2-s/2v$. 
This bound is always less than the limit of integration $t'$ in Eq.~(\ref{Iint}) because we always have
$s>v(t-t')$.}
\begin{equation}
I_{\cap}(t,t',s)=\frac{1}{4}\,I\left(\frac{t+t'}{2}\right)\exp\left({-\frac{\beta s}{2v}}\right)
\left[4+\frac{\beta s}{v}-\frac{[\beta (t-t')]^{2}}{\beta s/v}\right].
\label{Iintexp}
\end{equation}
The final integration with respect to $s$ in Eq.~(\ref{SSpriav}) must be done numerically.
In Fig.~\ref{figrmsexp} we show the surface correlation functions  
$\langle S(t)S(t')\rangle$, $\langle S(t)\rangle\langle S(t')\rangle$, 
and $\langle\Delta S(t)\Delta S(t')\rangle$.
The result depends on the nucleation time $t_N$, and we consider a bubble nucleated when $f_b=0.5$, i.e., at the time $t_\mathrm{med}=t_e+\beta^{-1}\log\log2$. 
The main difference with other nucleation times is the height of the curves, since older bubbles are larger 
(see the left panel of Fig.~\ref{figSexp}). Notice also that the area is measured in the natural units $(v/\beta)^2$; 
in units of the distance $d=v\Delta t$, the values are smaller.
The shapes of the equal-time curves (left panel) are qualitatively similar to those of the simultaneous nucleation case.
The main quantitative difference is that the value of $\langle S(t)\rangle^2$ is not so close to
$\langle S(t)^2\rangle$ in this case. The right panel shows the contour plots of $\langle\Delta S(t)\Delta S(t')\rangle$,
which are also qualitatively similar to the simultaneous case.

\subsection{Correlation between different parts of a bubble wall}

We shall now consider  
the probability for two points on a given bubble wall to be 
uncollided  at a given time $t$. 
We must take $t'=t$ in Eqs.~(\ref{Pcond}-\ref{VIsb}). 
In this case we have $s=R\sqrt{2(1-\cos\theta)}=2R\sin(\theta/2)$.
The conditional probability that a
point $p$ is uncollided assuming that another point $p'$
is uncollided is given by Eq.~(\ref{Pcond}).
For simultaneous nucleation, 
we have $I_{\cap}=n_{b}V_{\cap}$, with $V_{\cap}$ given by Eq.~(\ref{Vintsimulteqt}),
and we obtain 
\begin{equation}
P^S_{p|p'}(t,\theta)=\exp\left[-\frac{4\pi}{3}\left(\frac{t-t_{*}}{\Delta t_{b}}\right)^{3}\left(\frac{3}{2}\sin(\theta/2)-\frac{1}{2}\sin^{3}(\theta/2)\right)\right].
\label{pcondeqt}
\end{equation}
For the exponential case, $I_\cap$ is given by Eq.~(\ref{Iintexp}), and at equal times
we obtain
\begin{equation}
P^S_{p|p'}(t,t_N,\theta)=
\exp\left\{e^{\beta (t-t_*)}\left[e^{-\beta (t-t_N)\sin(\theta/2)}	
\left(1+\frac{1}{2}\beta (t-t_N)\sin(\theta/2)\right)-1\right]\right\}.
\end{equation}
The joint probability that both $p$ and $p'$ are in the false vacuum is given by
Eq.~(\ref{Pconj}),
$P^S_{p,p'}=e^{-I(t)+I(t_{N})}P_{p|p'}$ (for the simultaneous case, the prefactor is just $e^{-I(t)}$).

\begin{figure}[tbh]
\centering
\includegraphics[height=6.6cm]{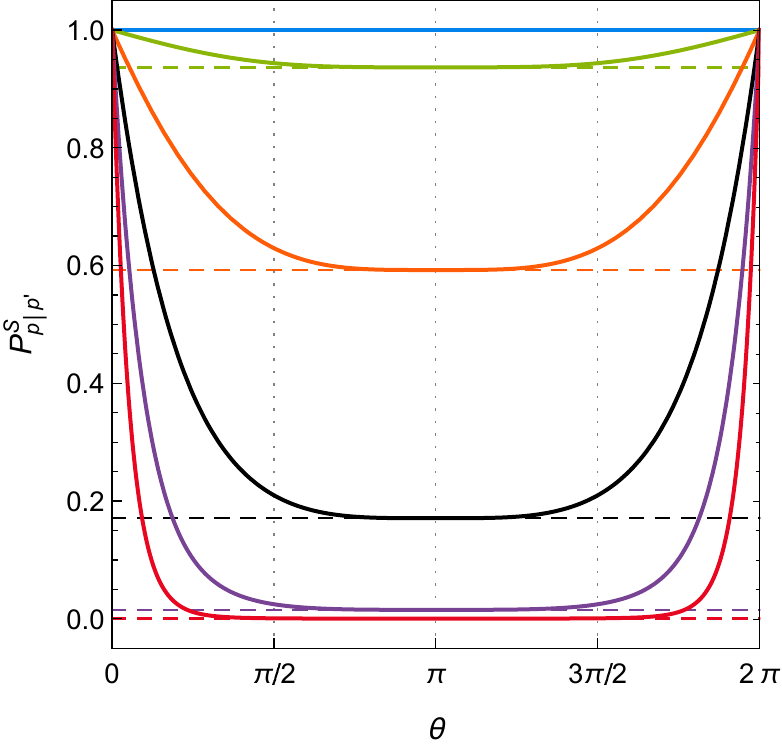} 
\hfill
\includegraphics[height=6.6cm]{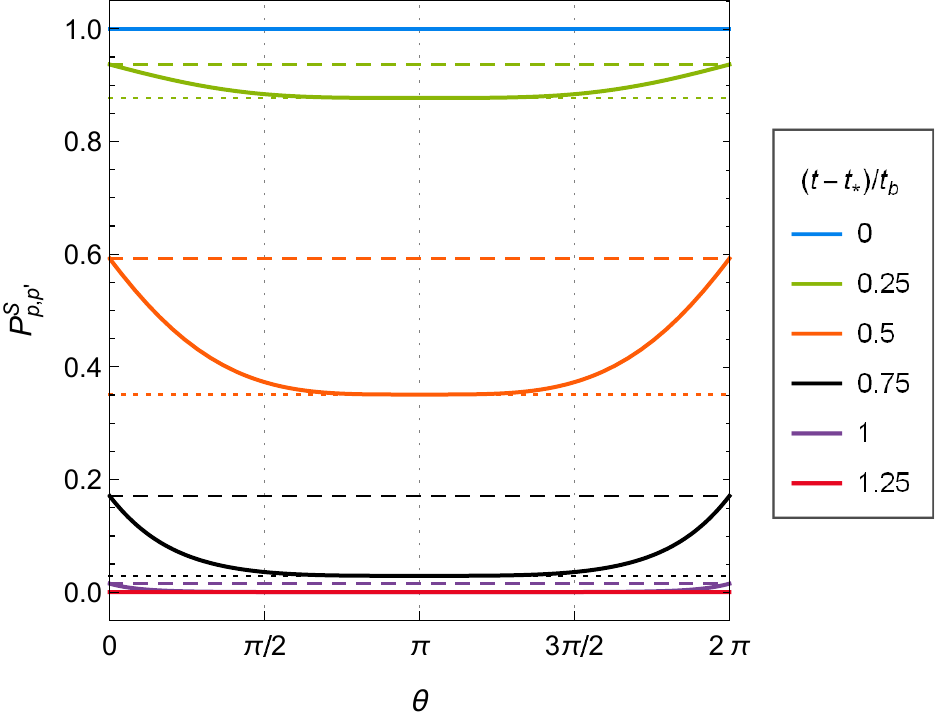}
\caption{Conditional probability (left) and joint probability (right) for two
points on a bubble wall to be uncollided in the case of simultaneous nucleation, 
as a function of the separation angle. The horizontal dashed lines indicate the value
of $P_{u}(t)=P_{fv}(t)$. The horizontal dotted lines indicate
the value $P_{u}(t)^{2}$. \label{fig2peqt}}
\end{figure}

In Fig.~\ref{fig2peqt} we show the result for simultaneous nucleation. 
At the beginning of the phase transition we have $P^S_{p,p'}=P^S_{p|p'}=1$, since
the two points are uncollided because the whole
bubble is isolated. By the end
of the phase transition, the probability that both points are uncollided
vanishes unless we assume that one of them is uncollided. In this
case (left panel), for $\theta\simeq0$ the probability will never vanish.
At intermediate times, assuming that $p'$ is uncollided, the probability
that $p$ is also uncollided falls with the distance from $p'$. 
For $\theta=\pi$  
Eq.~(\ref{pcondeqt}) gives the value $P_{u}(t)$, which is the probability
for an arbitrary point. 
This indicates that the correlation is lost exactly 
at the antipodal point. This happens because in this model all the bubbles have
the same radius $R$, and the correlation must vanish beyond a distance $2R$. 
Indeed, for both points $p$ and $p'$ to be affected by the same bubble, their separation 
must be smaller than $2R$. 
On the other hand, we see that the limit $P^S_{p|p'}=P_u$ is approached already
for $\theta\gtrsim \pi/2$.
The behavior of the joint probability is similar, except that it
takes the single-point value $P_{u}(t)$ for $\theta=0$, while for $\theta=\pi$ it takes the 
value $P_{u}(t)^2$, corresponding to independent variables.

\begin{figure}[tbh]
\centering
\includegraphics[width=0.485\textwidth]{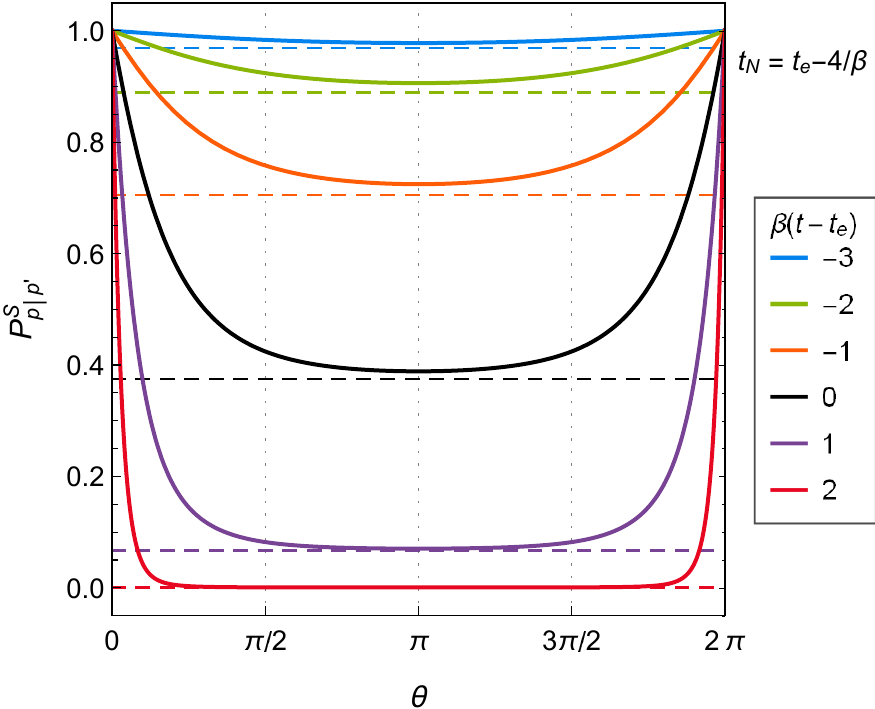} 
\hfill
\includegraphics[width=0.485\textwidth]{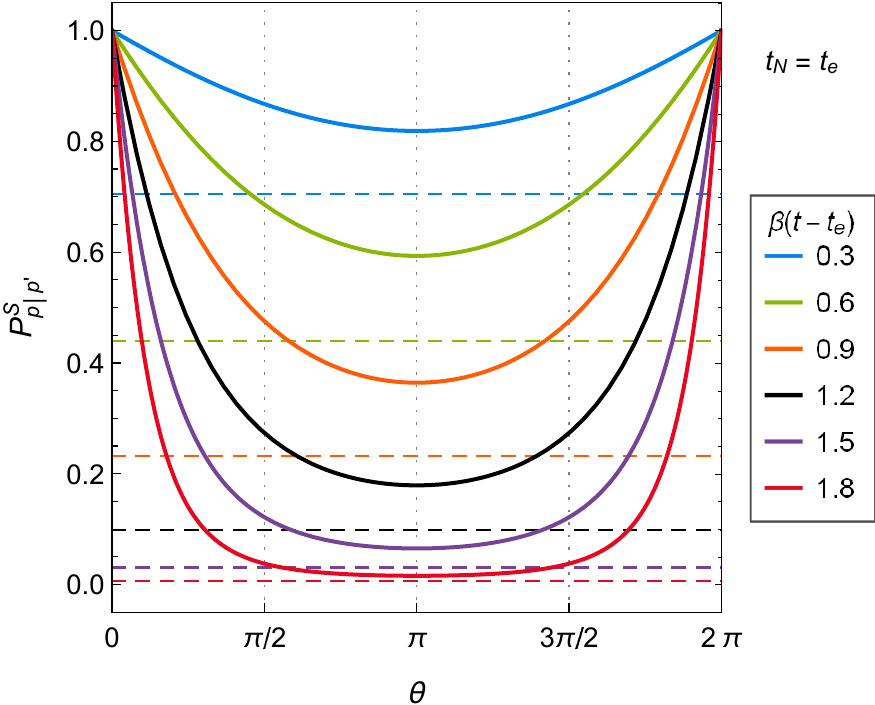}
\caption{Conditional probability as a function of the separation angle 
in the exponential case, 
at different times $t$ for a bubble nucleated at $t_N=t_e-4\beta^{-1}$ (left)  
and at $t_N=t_e$ (right).
The horizontal dashed lines indicate the value
of $P_u(t,t_N)=P_{fv}(t)/P_{fv}(t_N)$. \label{fig2peqtexp}}
\end{figure}

In Fig.~\ref{fig2peqtexp} we show the behavior of the conditional probability for the exponential case
(the joint probability has a simple relation with the latter, like in the simultaneous case). In the present case, 
$P^S_{p|p'}$ depends on the nucleation time $t_N$. 
We consider two different nucleation times $t_N$ in each panel of Fig.~\ref{fig2peqtexp},
and we show the probability $P^S_{p|p'}(\theta)$  
at different times $t$ like we did for the simultaneous case.
The main difference with that case is the fact that, at the maximum distance ($\theta=\pi$), 
the probability does not take
the single-point value $P_u(t,t_N)$ but a higher value, $P_u(t,t_N)\exp[I(t_N)\beta(t-t_N)/2]$.
This means that the point $p$ is never independent of the point $p'$: if $p'$ is uncollided, the 
probability of $p$ being also uncollided is always higher than without this condition. 
The value $P_u(t,t_N)$ can only be reached in the trivial limit $t\to t_N$ (i.e., 
if the bubble has just nucleated), and in the limit $t_N\to -\infty$. 

In the left panel of Fig.~\ref{fig2peqtexp} we considered a relatively early 
nucleation time, $t_N=t_e-4\beta^{-1}$, and times $t$ up to 
$t_e+2\beta^{-1}$ (at which the phase transition is already completed,
as can be seen, e.g., in Fig.~\ref{figSexp}).
We see that the result is qualitatively similar to the simultaneous case: 
although the probability never takes the single-point value $P_u(t,t_N)$, 
all the curves approach this value for $\theta\simeq \pi$. 
However, a bubble nucleated so early
is not representative, since bubbles which nucleate later are much more abundant.
In the right panel of Fig.~\ref{fig2peqtexp} we consider a bubble nucleated
at $t_N=t_e$,  at which the maximum nucleation of bubbles occurs, and
which is close to the time $t_\mathrm{med}$ at which $f_b=0.5$. 
In this case, the correlation between the points $p$ and $p'$ is never lost.
This happens because we have a wide range of bubble sizes.
Therefore, two points on the wall of a bubble of radius $R$ may be both affected by 
a single bubble of radius larger than $R$. 
Only in the case of a very large bubble (nucleated at $t_N\ll t_e$) this does not hold, 
since two points on its surface can only be affected by bubbles of smaller radii.  
In such a case the point correlation decreases significantly with the
separation, as observed in the left panel.

\section{Applications to cosmological consequences}
\label{cosmo}

We shall now discuss how some of the quantities derived in previous sections
enter the consequences of the phase transition.

\subsection{Electroweak baryogenesis}

In the electroweak phase transition,
the violation of baryon number takes place in the symmetric phase
outside the bubbles
and is biased by net charge densities generated in front of the moving walls.
The resulting baryon number density $n_{B}$ 
depends on 
the wall velocity. For very slow velocities, the plasma
will be
near equilibrium, and the net baryon number will vanish. On the other
hand, for very high velocities (close to the speed of sound), the
processes which violate baryon number (sphalerons) will have no time
to generate a significant amount of baryons as the wall passes. As
a consequence, there is a maximum baryon generation at a velocity
in the range $10^{-3}$-$10^{-1}$ (see \cite{ck20} for a recent
discussion). In most computations of electroweak baryogenesis for
specific models, the global dynamics of the phase transition is not
taken into account. In particular, the wall velocity is estimated
at the ``onset'' of nucleation (i.e., at the instant at which there
is a bubble per Hubble volume). However, for velocities in the above
range, which correspond to deflagrations, there will be a reheating
in the symmetric phase which will cause the wall velocity to decrease
from the initial value.

For the small velocities which are favorable for baryogenesis, a homogeneous
reheating can be assumed \cite{h95,ma05,m01}. 
The time-temperature relation for this case was derived
in Ref.~\cite{m04}. The expression for $T(t)$ 
contains a term which decreases with the scale factor 
(accounting for the adiabatic cooling)
and a term proportional to the fraction of volume occupied by bubbles, $f_{b}$
(accounting for the reheating). In many scenarios there is little supercooling,
$(T_{c}-T)\ll T_{c}$
\cite{ms08}, and we can linearize quantities which vanish at $T=T_{c}$,
except for those with a rapid variation, such as $\Gamma$ or $f_{b}$. 
Thus, e.g., the adiabatic cooling relation becomes $T_{c}-T=T_{c}H(t-t_{c})$,
but the release of latent heat introduces a term proportional to $Lf_{b}(t)$,
where $L$ is the latent heat.
With these approximations, analytic expressions
for the development of the phase transition 
were obtained in Ref.~\cite{mr18}. Due to the high
sensitivity of the nucleation rate with the temperature, a simultaneous
nucleation at a certain time $t_{*}$ (corresponding to the minimum
temperature reached $T_{*}$) is a good approximation.
The subsequent evolution
depends on the parameter
\begin{equation}
q\simeq\frac{L}{\rho_{R}(T_{c})-\rho_{R}(T_{*})},\label{q}
\end{equation}
where $\rho_{R}(T)$ is the energy density
in radiation. Thus, $q$ gives the ratio of the released energy to
the energy which is needed to reheat the plasma back to $T_{c}$.
This parameter is important since the wall velocity vanishes at $T=T_{c}$.
In particular, the approximation  
$v\propto T_{c}-T$ leads to
\begin{equation}
\frac{v}{v_{*}}=1-qf_{b}(t)+\frac{t-t_{*}}{t_{*}-t_{c}}.
\end{equation}
Since the nucleation is simultaneous, we have $f_{b}=1-e^{-I}$, with $I=\frac{4\pi}{3}(R/d_{b})^{3}$.

There are two well differentiated behaviors depending on
whether the value of $q$ is greater or less than 1 \cite{mr18}. For $q<1$ the
function $R(t)$ can be approximated by the relation 
\begin{equation}
\frac{t-t_{*}}{d_{b}/v_{*}}=\frac{R}{d_{b}}+
\frac{\pi q}{3}\left(\frac{R}{d_{b}}\right)^{4}.\label{tR}
\end{equation}
On the other hand, for $q>1$, the reheating takes the temperature
very close $T_{c}$, and the wall velocity may decrease by a few orders
of magnitude. As a consequence, after an initial reheating stage in
which the approximation (\ref{tR}) is valid, a relatively long stage
of very slow growth is established. Assuming an approximate phase
equilibrium during this second stage, the evolution can be obtained
by equating the released energy density $L\langle S_{\mathrm{tot}}\rangle
vdt$
to the decrease of energy density due to the expansion, $4\rho_{R}Hdt$
(since $\rho_{R}\propto T^{4}$ and $\dot{T}=-HT$). This gives 
\begin{equation}
\frac{v}{v_{*}}=\frac{d_{b}/v_{*}}{t_{*}-t_{c}}\frac{e^{I}}{q4\pi(R/d_{b})^{2}}.
\end{equation}
Here, the relations $\langle S_{\mathrm{tot}}\rangle=\langle S\rangle n_{b}$,
$\rho_{R}(T_{c})-\rho_{R}(T_{*})\simeq4\rho_{R}(T_{c}-T_{*})/T_{c}$,
and $(T_{c}-T_{*})=T_{c}H(t_{*}-t_{c})$ have been used. An approximation
for the relation $R(t)$ in this slow stage is given in \cite{mr18},
\begin{equation}
\frac{t-t_{*}}{d_{b}/v_{*}}=\frac{t_{*}-t_{c}}{d_{b}/v_{*}}(qf_{b}-1)+\frac{e^{I}}{q4\pi(R/d_{b})^{2}}.
\end{equation}

\begin{figure}[tbh]
\centering
\includegraphics[width=0.49\textwidth]{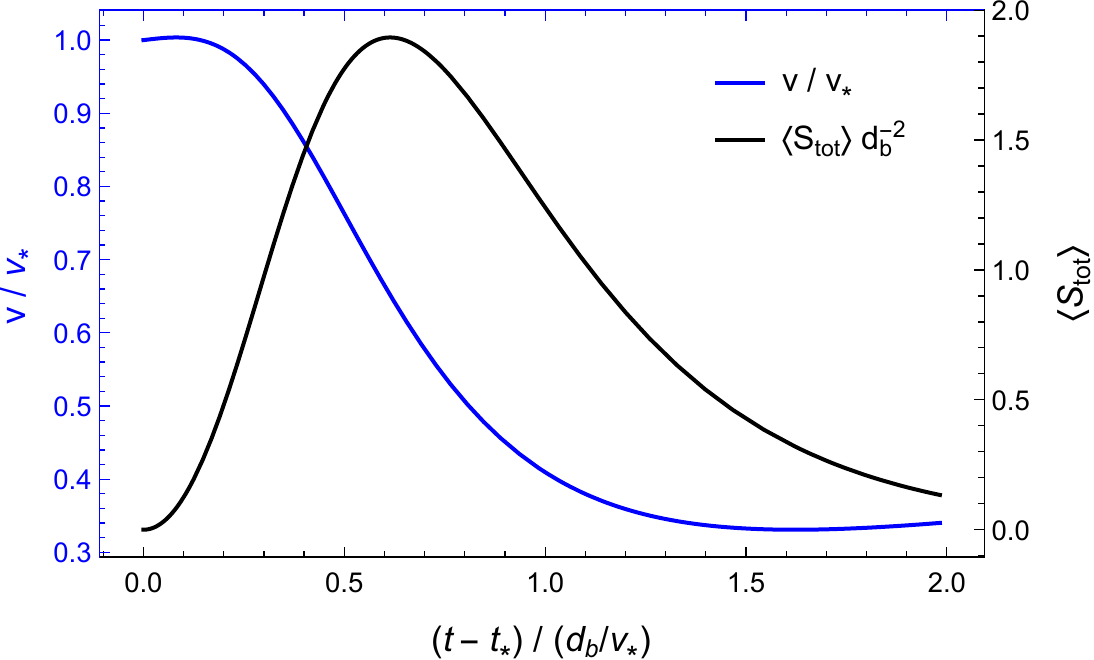}
\hfill
\includegraphics[width=0.49\textwidth]{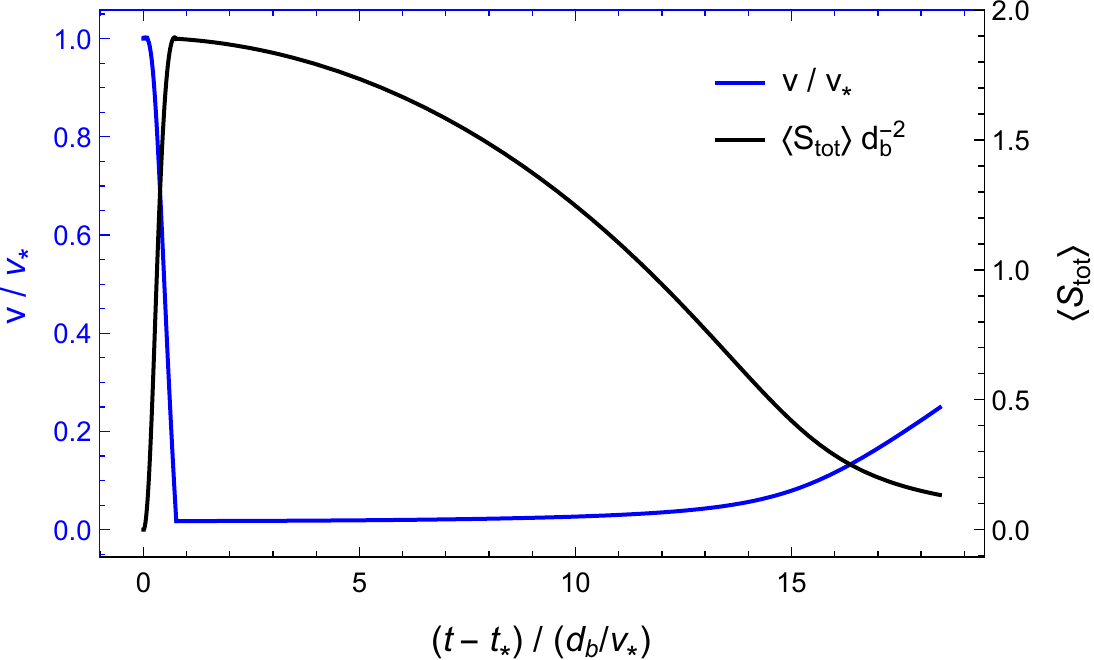}
\caption{The wall velocity and the average wall area for $q=0.8$ (left) and
$q=2$ (right).\label{figvelo}}
\end{figure}

The evolution of the wall velocity for these approximations
is shown in Fig.~\ref{figvelo} for the two 
cases $q<1$ and $q>1$, together with the average uncollided
surface. For the case $q<1$ we considered a relatively high value $q=0.8$, so that we obtain a velocity variation of order 1.
The time $t_{*}-t_{c}$ depends on the supercooling 
parameter $(T_{c}-T_{*})/T_{c}$.
This time is usually an order of magnitude greater than the time scale
of the reheating, which is given by $d_{b}/v_{*}$. Therefore, we
used the value $(t-t_{*})/(d_{b}/v_{*})=15$. For the case $q>1$
we considered the same supercooling parameter, and we chose $q=2$.
For $q\gg1$ the effect will be more pronounced. 

The velocity decrease may suppress or enhance the baryon generation,
depending on whether the initial velocity $v_{*}$ is smaller or higher
than the maximum of $n_{B}$ \cite{h95,ma05,m01}. If the velocity
variation occurs around the maximum, then it will not cause a significant
effect on baryogenesis. Therefore, we shall consider only the cases in
which the velocity variation occurs entirely on the left or on the
right of the maximum of $n_{B}(v)$. For simplicity, we shall assume
that in these cases the dependence is either $n_{B}\propto v$ or
$n_{B}\propto v^{-1}$ (for a motivation of this dependence and analytic
approximations for the whole velocity range, see \cite{m01,ma05}).
If the baryon number density generated at the wall is of the form
$n_{B}\propto v$, the total baryon number produced will be suppressed
with respect to the initial value $n_{B*}$. In Fig.~\ref{figvelo} 
we see that the average
wall area, which acts as a weight, emphasizes this effect, since 
$\langle S_{\mathrm{tot}}\rangle$
is maximal when the baryon density has already begun to decrease.

\begin{figure}[tbh]
\begin{centering}
\includegraphics[width=0.49\textwidth]{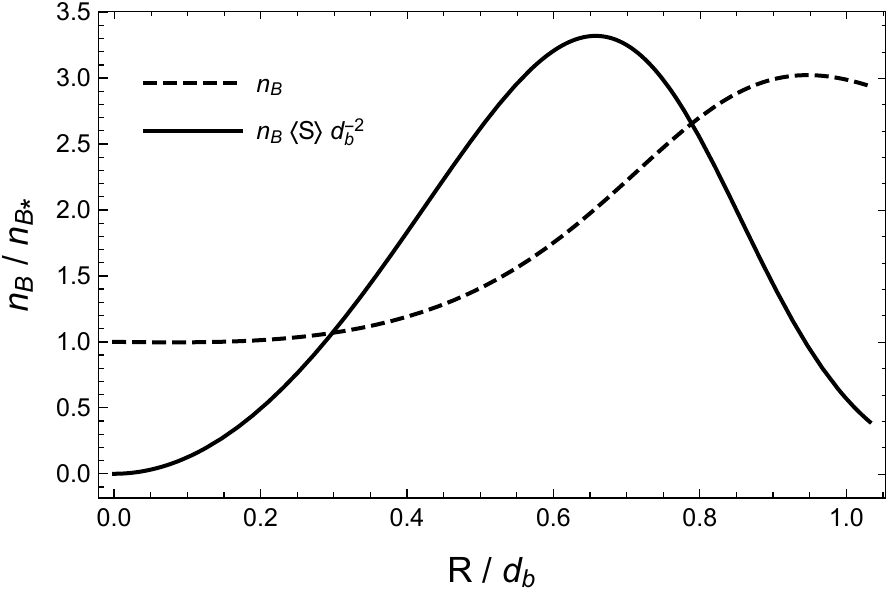} 
\hfill
\includegraphics[width=0.49\textwidth]{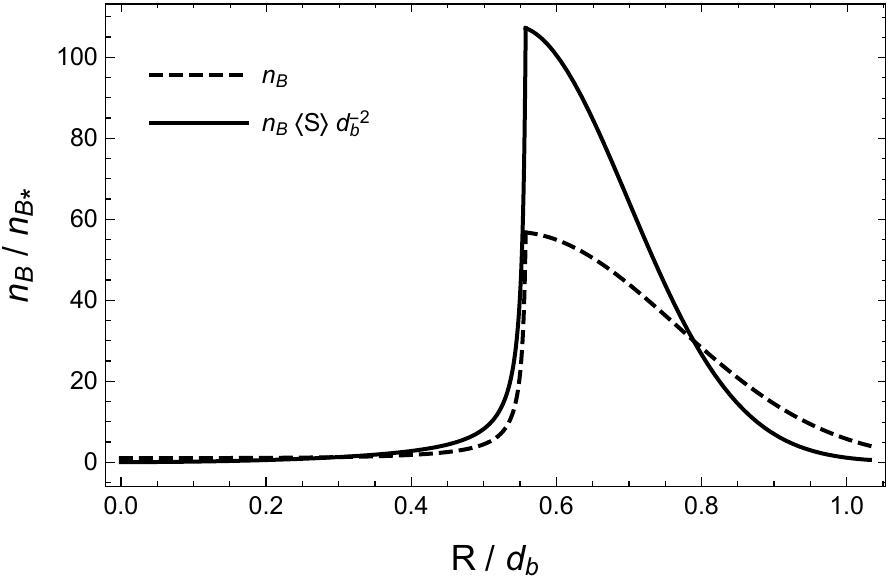}
\caption{The baryon number density as a function of the bubble radius for the
cases of Fig.~\ref{figvelo}. 
\label{figperfil}}
\par\end{centering}
\end{figure}

In the case $n_{B}\propto v^{-1}$ the total baryon density is enhanced
with respect to $n_{B*}$. Let us consider this case in more detail,
since it is more interesting. We may write $n_{B}(v)=n_{B*}v_{*}/v.$
The total baryon number produced at time $t$ in a volume $V$ is
given by
\begin{equation}
VdB=n_{B}(v(t))\langle S_{\mathrm{tot}}(t)\rangle v(t)dt.
\end{equation}
The expression is more clear in terms of the bubble radius,
\begin{equation}
dB=n_{B}(v)\langle S\rangle dR/d_{b}^{3},
\end{equation}
with $\langle S\rangle=4\pi R^{2}e^{-\frac{4\pi}{3}(R/d_{b})^{3}}$.
Thus, in units of $d_{b}$, we have $\int_{0}^{\infty}\langle S\rangle dR=1$.
In Fig.~\ref{figperfil} we show the local baryon number density
$n_{B}$ as a function of $R$, together with the weighted density
$n_{B}\langle S\rangle$. The function $n_{b}(R)$ gives the profile
of the baryon inhomogeneities produced inside each bubble, which may also
be of interest, while the product $n_{B}\langle S\rangle$ gives, upon integration, the
average baryon number density $B$. 
For $q<1$, the weight function peaks somewhere between the minimum
and maximum value of $n_{B}$. On the other hand, for $q>1$, the
highest values of $n_{B}$ have a higher weight. This can be inferred
already from Fig.~\ref{figvelo}. In these specific examples, the total amplification
is $B/n_{B*}\simeq1.74$ for the case $q=0.8$ and $B/n_{B*}=21.2$
for the case $q=2$.

\subsection{Gravitational waves}

As mentioned in the introduction, the moving walls cause different
phenomena  which generate GWs, such as sound waves and turbulence, and
the walls themselves are a direct source of gravitational radiation. Besides
the well known bubble-collision mechanism, 
other kinds of deformations from the spherical shape may produce GWs. 
The exponential growth
of wall corrugations due to hydrodynamic instabilities \cite{link,hkllm93,mm14,mms15}
may constitute an important example.
The relevant quantity in the calculation of gravitational waves is the  
transverse and traceless projection of the energy-momentum
tensor of the source, $ \Pi_{ij}(t,\mathbf{k})$ in Fourier space. 
The energy density per logarithmic wavenumber of GWs
can be written in the form \cite{cds08,jt17}
\begin{equation}
\frac{d\rho_{GW}}{d\ln k}=\frac{2Gk^{3}}{\pi}
\int_{-\infty}^{\infty}dt\int_{-\infty}^{\infty}dt'e^{ik(t-t')}\Pi(t,t',k),
\label{rhogw}
\end{equation}
where the quantity $\Pi$ is essentially the unequal-time correlator
$\left\langle \Pi_{ij}\Pi_{ij}^{*}\right\rangle $, after subtracting a 
delta function of the wave vector.
Below we shall relate the general characteristics of the spectrum of GWs originated at the bubble walls with the
average wall area and the its correlations.

Some general properties of the GW spectrum from phase transitions
were discussed in Ref.~\cite{cdks09}. 
For colliding bubbles, 
a statistical argument was used to motivate a source correlator of the form 
(with separated variables $t,t'$)
\begin{equation}
\Pi(t,t',k)=\frac{2}{(2\pi)^{3}}\rho_{s}^{2}\frac{N}{V}f(k,t)f(k,t'),
\label{Pestad}
\end{equation}
where the constant $\rho_{s}$ is the energy density of the source and
$N/V$ is the
density of collision events. This gives, according to Eq.~(\ref{rhogw}),
\begin{equation}
\frac{d\rho_{GW}}{d\ln k}=\frac{4G\rho_{s}^{2}}{\pi(2\pi)^{3}}\frac{N}{V}k^{3}
\left|\int_{-\infty}^{\infty}dte^{ik t}f(k,t)\right|^{2}\equiv Ck^{3}
\left|\hat{f}(k,k)\right|^2.\label{rhoellas}
\end{equation}
In Ref.~\cite{cdks09}, the function $f(k,t)$ was further assumed
to be of the form 
\begin{equation}
f(k,t)=g(t)F(k),\label{separ}
\end{equation}
where the function $g(t)$ vanishes outside a time interval $[t_{*},t_{*}+\Delta t]$
in which the phase transition occurs. Here, $t_{*}$ is the initial
time and $\Delta t$ is the duration of the phase transition. They
considered several examples for this function. 
For the spatial Fourier transform $F(k)$,
a simple function involving a characteristic
length scale $L$ of the problem was proposed. 
This function was constructed from some physical requirements and
according to the modeling of a previous analytic calculation \cite{cds08}. 
The separability assumption of Eq.~(\ref{separ}) 
does not reproduce the features found in the simulations \cite{hk08},
so a time-dependent variable $L(t)$ was considered.
The bubble radius $v(t-t_*)$ does not work either, and
$L(t)$ is argued to
represent
the size of a typical colliding region, which initially grows
but vanishes at the end of the transition. Therefore, $L(t)$
is modeled with 
\begin{equation}
L(t)=v\Delta t\,g(t),\label{Lg}
\end{equation}
and the final form of $f$ is
\begin{equation}
f(k,t)=L\sqrt{\Delta L}
\left[\frac{1+\left(\frac{k L}{3}\right)^{2}}{1+\left(\frac{k L}{2}\right)^{2}+
\left(\frac{k L}{3}\right)^{6}}\right]^{1/2}.
\label{fomt}
\end{equation}
The global square root is motivated by the coherent-source approximation
\begin{equation}
\Pi(t,t',k)=\sqrt{\Pi(t,t,k)}\sqrt{\Pi(t',t',k)},\label{coher}
\end{equation}
and the prefactor $\sqrt{L^{2}\Delta L}$ in Eq.~(\ref{fomt}) represents
the volume of an uncollided wall, where $\Delta L$ 
parameterizes the wall thickness. 

The result of the Fourier transform in Eq.~(\ref{rhoellas}) strongly
depends on the continuity properties of the function $g(t)$, and
the different functions considered in Ref.~\cite{cdks09} give quite
different behaviors of the GW spectrum. Based on results of computations
of two bubble collisions from Ref.~\cite{kt93}, it was argued that
the time dependence is continuous but not differentiable. Indeed,
using the simple function 
\begin{equation}
g(t)=\frac{4(t-t_{*})\left[\Delta t-(t-t_{*})\right]}{(\Delta t)^{2}},\label{g2}
\end{equation}
the construction (\ref{fomt})
gives a GW spectrum of the required form. In particular, it reproduces
the $k^{-1}$ decay for large frequencies found in simulations.

\begin{figure}[tbh]
\centering
\includegraphics[width=0.45\textwidth]{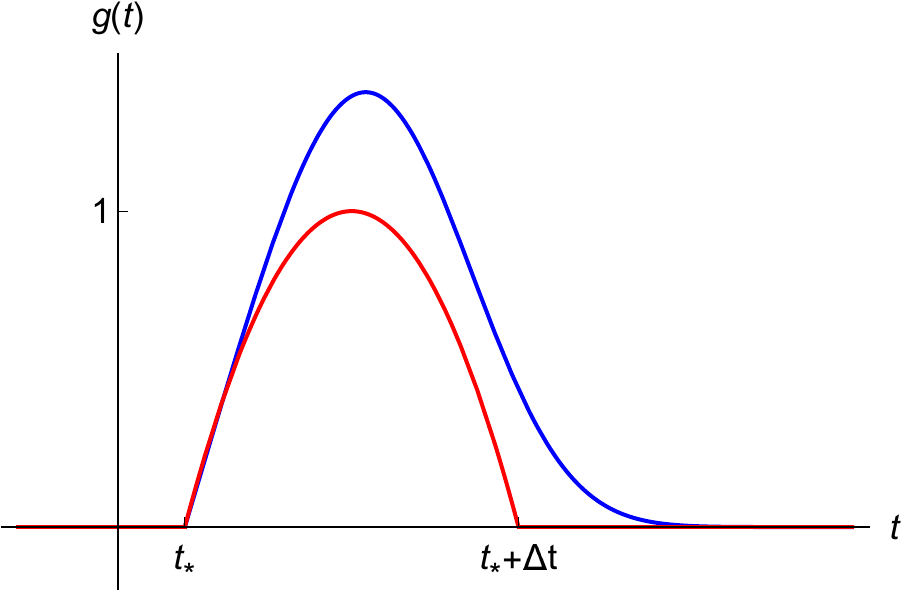}
\hfill
\includegraphics[width=0.50\textwidth]{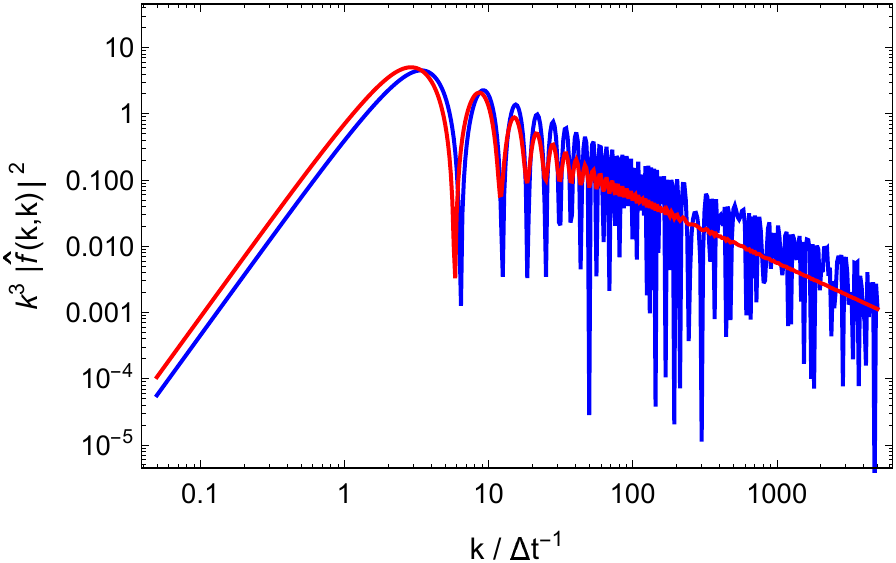}
\caption{Left: Time evolution of the length scale of colliding walls $g(t)$
given by Eq.~(\ref{g2}) (blue) and Eq.~(\ref{g}) (red). Right:
The corresponding spectra with the approximation (\ref{fomt}). \label{figg}}
\end{figure}

In our statistical treatment of the phase transition, the volume of
uncollided walls is proportional to their uncollided area. 
Therefore, we may write the quantity 
${L^{2}\Delta L}$ in the form ${\langle S\rangle\Delta L}$, and the size of the typical 
colliding region becomes
\begin{equation}
L(t)=\sqrt{\langle S(t)\rangle}.\label{LS}
\end{equation}
The average area is given by Eq.~(\ref{Sumed}), which roughly
grows quadratically with the bubble size at the beginning of the phase
transition and falls exponentially at the end. As we have seen, the
behavior
is qualitatively similar
for an exponential nucleation rate and for a simultaneous
nucleation. 
The latter is more consistent 
with the approximations of Eqs.~(\ref{Pestad}), (\ref{Lg}) and (\ref{fomt}) 
(in particular, the assumptions of a constant  $N/V$ and  
a single length scale $L$). Therefore, we shall consider this simpler case.
Comparing Eq~(\ref{LS}) with Eq.~(\ref{Lg}), we see that our definition of the function 
$g(t)$ should be\footnote{As we have seen, for simultaneous nucleation we have 
$\Delta t\simeq t_b$ and $v\Delta t\simeq d_b$.} 
$g=\sqrt{\langle S\rangle}/d_{b}$, i.e.,
\begin{equation}
g(t)=\sqrt{4\pi}\left(\frac{t-t_{*}}{\Delta t}\right)\exp\left[-\frac{2\pi}{3}\left(\frac{t-t_{*}}{\Delta t}\right)^{3}\right]\label{g}
\end{equation}
for $t>t_{*}$, and $g(t)=0$ for $t<t_{*}$. In Fig.~\ref{figg}
we compare these functions and the corresponding form of GW spectra. These
spectra have the same asymptotic behavior (namely, proportional to
$k^{3}$ at low frequencies and to $k^{-1}$ at high frequencies).
They have also a very similar shape near the peak, but they depart quantitatively
for $k\gtrsim10\Delta t^{-1}$. 

\begin{figure}[tbh]
\centering
\includegraphics[width=0.49\textwidth]{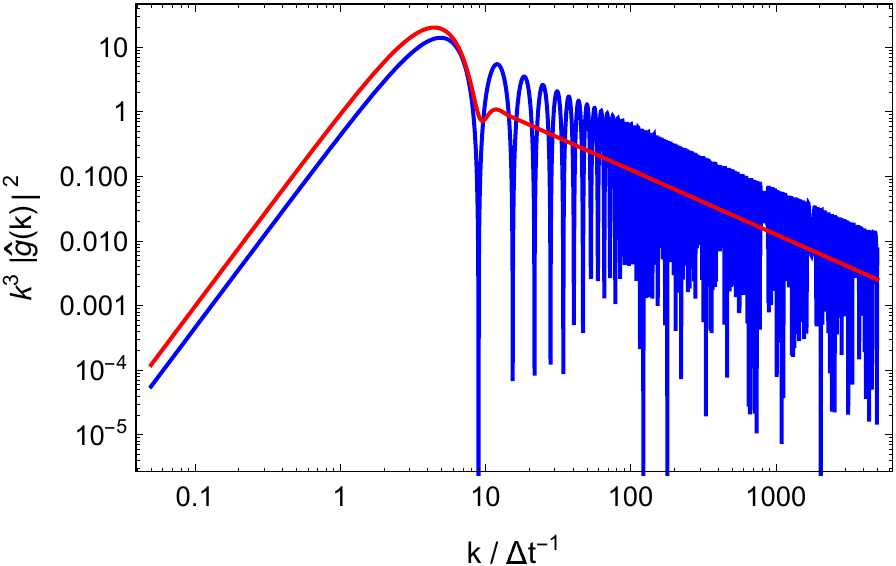}
\hfill
\includegraphics[width=0.49\textwidth]{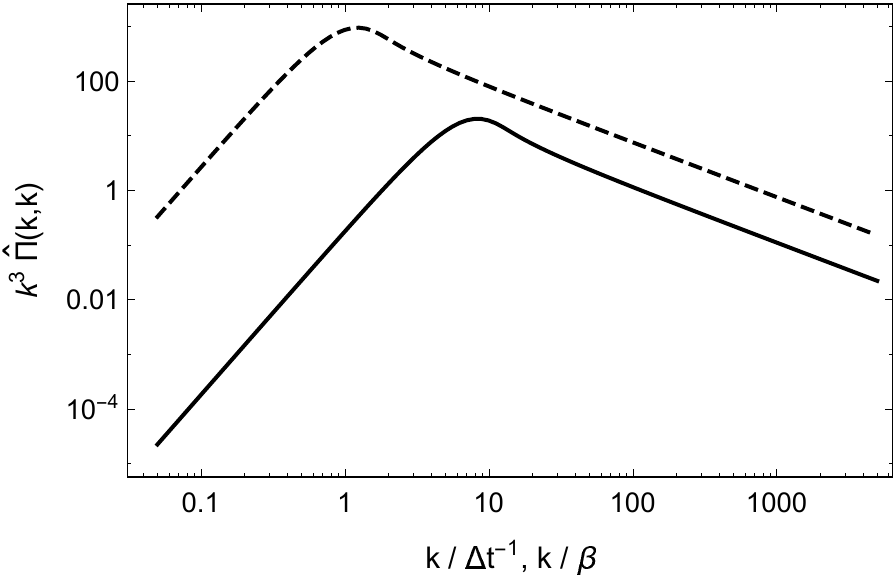}
\caption{Spectrum obtained without assuming a particular dependence on $k L$. 
Left: Like in Fig.~(\ref{figg}), but using the ``window'' function $g(t)$ alone.
Right:  Assuming a source proportional to $\langle\Delta S(t)\Delta S(t')\rangle$, with 
time units $\Delta t$ (solid line) and $\beta^{-1}$ (dashed line). \label{figspec0}}
\end{figure}

We remark that, in contrast to the quadratic function 
(\ref{g2}), our function (\ref{g}) was derived from the average
surface $\langle S\rangle$, which gives the actual ``time window''
inside which the source is active. This window will be present for
any phenomenon originated at the bubble walls, such as GWs from corrugation
instabilities. 
In order to investigate how the general features of the GW spectrum
depend on this function and to what extent
they depend on the spatial Fourier transform,
let us set $F(k)=$
constant in Eq.~(\ref{separ}). Thus, Eq.~(\ref{fomt}) becomes 
$f(k,t)=L(t)\sqrt{\Delta L}$, and
we obtain a GW spectrum proportional to $k^{3}|\hat{g}(k)|^2$.
The form of this spectrum is shown in Fig.~\ref{figspec0}, where
we also consider the approximation (\ref{g2}) for $g(t)$.
We see that the function $g(t)$ alone reproduces the 
general characteristics of the spectrum.
Leaving
aside the fact that this approximation gives only the schematic form
of the spectrum, we notice that the definition of $g(t)$ from the
average uncollided surface gives a more physical result. In particular,
the quadratic approximation for $g(t)$ produces spurious oscillations
in which the spectrum vanishes periodically. 

The rough approximation we obtained by combining Eq.~(\ref{LS}) with the previous
approximations corresponds to using the coherent-source
approximation (\ref{coher}), with an equal-time correlator proportional
to $\langle S\rangle(t)$,
\begin{equation}
\Pi(t,t')=\frac{2\rho_{s}^{2}\Delta LN/V}{(2\pi)^{3}}
\sqrt{\frac{\langle S(t)\rangle\langle S(t')\rangle}{d_b^4}}.
\label{raiz}
\end{equation}
As already mentioned, it is to be expected that the relevant quantity
is the correlation 
function\footnote{In this approximation it makes no sense to consider the two-bubble correlation function $\langle S(t)S'(t')\rangle$.}
$\langle S(t)S(t')\rangle$ rather than
$\langle S(t)\rangle\langle S(t')\rangle$. 
Nevertheless, these two quantities are not too different,
as we have seen in the previous section (see Fig.~\ref{figrms}).
If we replace $\langle S(t)\rangle\langle S(t')\rangle$ with $\langle S(t)S(t')\rangle$, there is no actual reason 
for keeping the square root in Eq.~(\ref{raiz}), which was motivated by the coherent approximation.
In any case, the square root does not change the qualitative features of the resulting spectrum, and 
a source of the form $\Pi(t,t')\propto\langle S(t) S(t')\rangle$ seems more reasonable. 
Moreover, these rough approximations
do not take into account the transverse-traceless projection of the energy-momentum tensor, 
which, in particular, prevents a spherical bubble to radiate.
The result should actually
vanish in the uncorrelated case $\langle S(t) S(t')\rangle=\langle S(t)\rangle\langle S(t')\rangle$.
To take this into account, we shall consider the approximation 
\begin{equation}
\Pi(t,t')\propto\langle\Delta S(t)\Delta S(t')\rangle. 
\end{equation}
In the right panel of Fig.~\ref{figspec0} we show the result of the Fourier transform for this 
model, 
\begin{equation}
\hat{\Pi}(k,k)=\int dt\int dt'e^{ik(t-t')}\Pi(t,t').
\end{equation} 
The solid line corresponds to the same time units considered in the other figures, namely, 
the time interval $\Delta t$ in which the source is active. 
This curve has the same general characteristics as the others, but its shape is in agreement with more rigorous computations
\cite{hk08,jt17}. 
The dashed line is obtained by converting to the time unit $\beta^{-1}$ 
for an exponential nucleation rate, which for this model is rather artificial 
and is defined through the  
relation $\beta^{-1}\simeq \Delta t/6.13$ (see Sec.~\ref{Smedia}).
We see that the peak of the spectrum is at $k/\beta \simeq 1$, in agreement with  \cite{hk08,jt17}
(the amplitude of the spectrum does not have a physical meaning since we have omitted any proportionality constants).

\subsection{Topological defects}

The probability for different points on a bubble wall to be 
collided is a basic ingredient in the calculation of the
dynamics of bubble intersections which enter the mechanism of defect
trapping. 
This probability,
which we have studied in previous sections for two points, 
is only the first step towards a calculation of 
defect formation using the statistical method considered in this paper. 
The probability of trapping a vortex, as explained in Sec.~\ref{intro}, 
is related to that for three or more bubbles 
to get in contact with each other and enclose a region of false vacuum. 
Therefore, each of these bubbles must have at least two separate collided regions.
It is clear that our method can be generalized to address
this calculation. However, combining these probabilities 
with the condition that the overlapping bubbles form a closed chain
does not seem trivial.
Notice that the probability of forming a chain of several overlapped bubbles 
will also provide an analytic approach to the study of bubble percolation.
We shall consider these applications elsewhere.

Nevertheless, from the two-point probabilities 
we can already see some differences between models
which will be relevant for defect formation.
An important element of the dynamics of defect formation is
the time it takes a third bubble to arrive once two bubbles have collided.
Therefore, the correlation between different parts of a bubble wall is very important.
As we have seen in Sec.~\ref{somecases}, 
for a simultaneous nucleation the point correlation 
falls relatively quickly with the distance between the points, 
while for an exponentially growing rate all the points on the surface remain correlated.
This implies that in the former case it will be more probable that a given bubble 
at a given time has collided with more than one bubble.
From the probabilities $P^S_{p|p'}$ and $P^S_{p,p'}$ derived in Sec.~\ref{pointsonbubbles}
we may obtain  two-point 
probabilities which are more directly related to such multiple collisions.
For instance, 
the probability that a point $p$ is collided 
assuming that another point $p'$ on the same bubble wall is collided
is  given by $(1-2P_u+P^S_{p,p'})/(1-P_u)$, while 
the probability that $p$ is collided assuming that $p'$ is not collided 
is given by $1-P^S_{p|p'}$.
The independent value of both probabilities is $1-P_u$.

\begin{figure}[tbh]
\centering
\includegraphics[height=5.3cm]{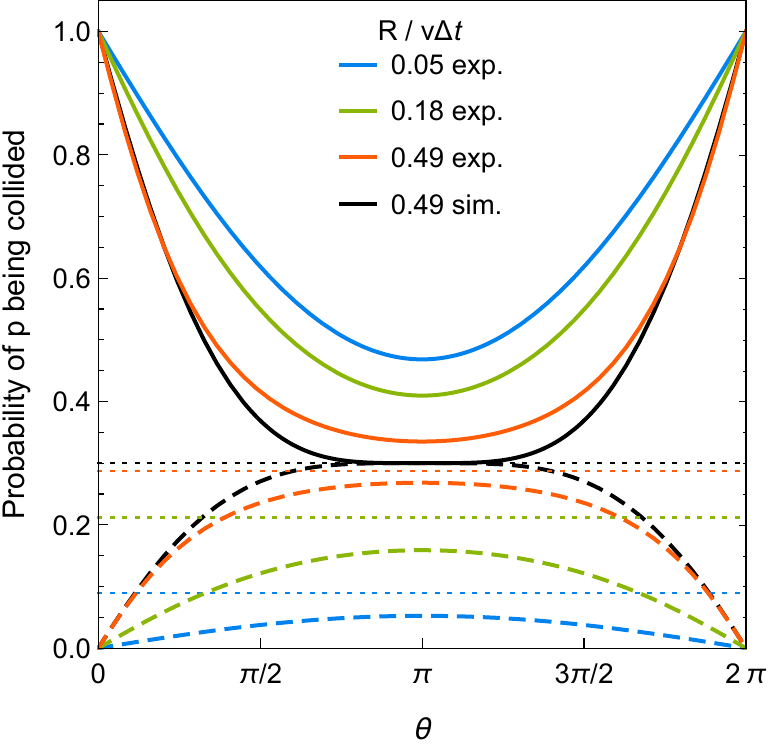} 
\hfill
\includegraphics[height=5.3cm]{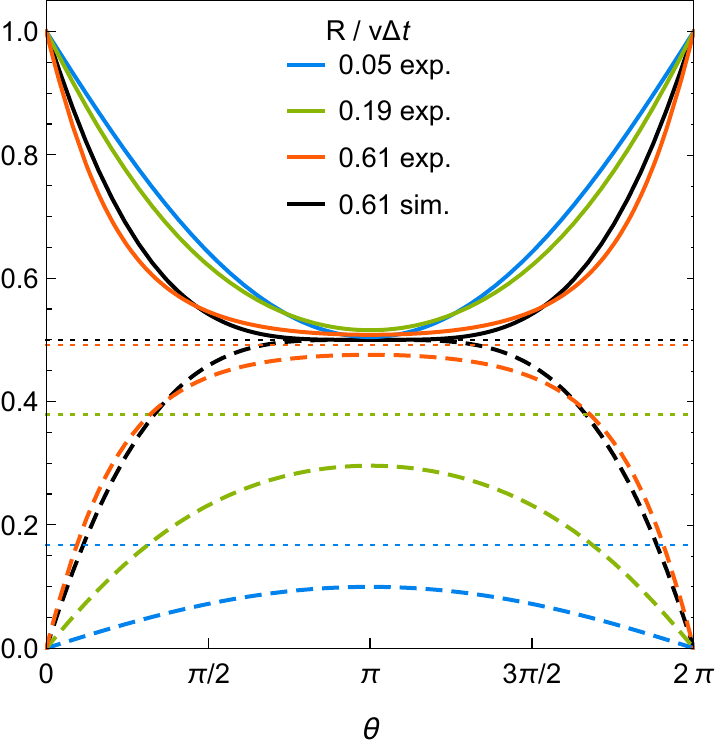}
\hfill
\includegraphics[height=5.3cm]{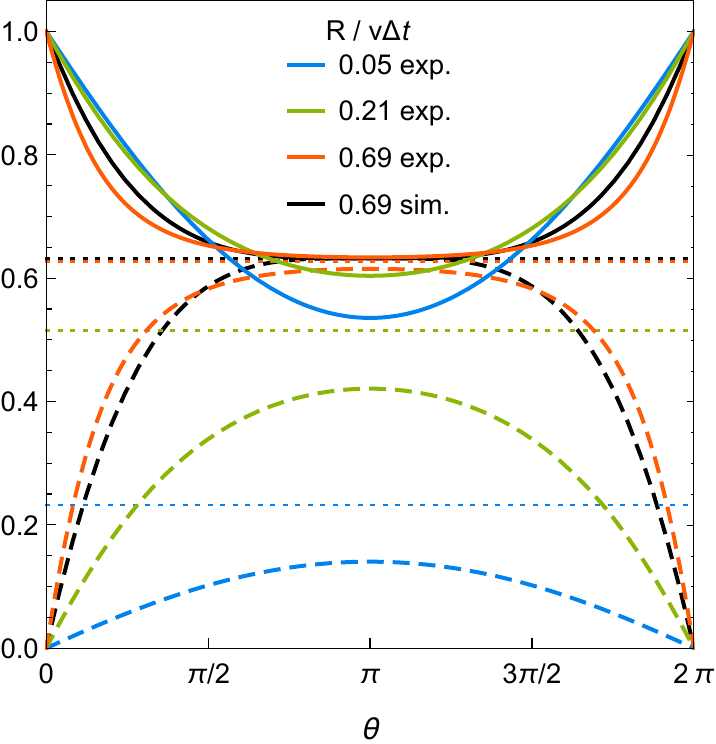}
\caption{Probability that $p$ is collided given that $p'$ is collided (solid lines) 
and given that $p'$ is not (dashed lines),
when $f_b=0.3$ (left), $f_b=0.5$ (center), and $f_b\simeq 0.63$ (right).
The horizontal dotted lines indicate the values
of $1-P_u(t,t_N)$.
The black line corresponds to the delta-function nucleation rate and the others to
the exponential nucleation rate.
 \label{fig2pr}}
\end{figure}

In Fig.~\ref{fig2pr} we consider these probabilities
for our two nucleation models.
Each line corresponds to a given bubble radius and the probabilities are plotted 
as a function of the angle between the points.
For a better comparison of the models, 
we use the unit of length $v\Delta t$ given by the duration of the phase transition, and
the three panels show the situation at three representative moments
corresponding to given values of the volume fraction.
From left to right, we considered 
the time at which $f_b=0.3$ (this gives a rough estimation of the percolation 
time\footnote{Simulations with overlapping spheres of equal radius give $I\simeq 0.34$ and 
$f_b\simeq 0.29$  (see, e.g., \cite{Rintoul_1997}).}),
the time at which $f_b=0.5$, and the time at which $f_b=1-e^{-1}$.
The black lines correspond to the simultaneous nucleation case, for which 
there is a single bubble size at a given time. For the case of the exponential rate,
we consider three bubble sizes. One of them has the same value of the simultaneous case
(orange lines), and the other two are smaller. 

In the simultaneous case, 
the probability that a point is collided always approaches the independent value $f_b(t)$
for an angle  $\theta\approx \pi/2$. 
In the exponential case, for a bubble of the same radius
the behavior is quite similar. 
However, 
for an exponential rate those bubbles are not relevant since smaller bubbles are much
more abundant. 
The green lines correspond to the average bubble radius at each time and
the blue lines correspond to an even smaller bubble.
For these  bubbles the behavior is quite different.
In the first place, the general probability that a point $p$ on the bubble wall is collided
(dotted lines) decreases 
from  the value $f_b$ corresponding to an arbitrary point in space (black dotted line).
If we further assume that another point $p'$ on the wall is uncollided, 
the probability that $p$ is collided is even smaller (dashed lines). 
The separation between dotted and dashed lines indicates that the probabilities 
for the two points  are never independent. 

On the other hand, if we assume that a point $p'$ on the wall is already collided,  
the probability that $p$ is collided is higher (solid lines).  
For $\theta=0$ this probability is 1, while  
for $\theta=\pi$ it gets more or less close to the value $f_b$, 
depending on the bubble size and the time.
However, it is important to remark that this is not the independent value
for a point on the surface but for an arbitrary point in space.
For small bubbles, the solid lines are far from the
dotted lines, which means that the probability of $p$ being collided
is quite higher than the independent value if $p'$ is collided.
Therefore, under this condition, $p$ has a probability $\sim f_b$ of being collided, 
but it is likely that $p$ has collided with \emph{the same} bubble as $p'$.
In contrast, 
for the simultaneous case the solid curve rapidly approaches the independent value, 
indicating that we have a probability 
$f_b$ that the point has collided with \emph{any} bubble.
In this scenario, simultaneous multiple collisions,
which favor defect trapping, are more probable.

\section{Conclusions}

\label{conclu}

We have studied bubble wall correlations in cosmological phase transitions using 
a statistical treatment of the bubble kinematics.
Specifically, we have calculated conditional and joint probabilities for a set of
arbitrary points of space to remain in the false vacuum at different
times, conditional and joint probabilities for two points on a bubble wall 
or on two different
bubble walls to remain uncollided, and the probability that 
a point on a bubble wall is uncollided, in the presence of another bubble. 
We have used these probabilities to study the evolution of the envelope of uncollided walls
as well as spacial and temporal correlations within this surface.
Our general results depend on the nucleation rate $\Gamma(t)$ and the wall velocity
$v(t)$, and can thus be applied to different types of phase transitions.
We have considered a few specific models and discussed the application 
of our results 
to the calculation of possible remnants of the phase transition, such as 
the baryon asymmetry of the Universe,
a stochastic background of gravitational 
waves, and topological defects.

Our general treatment of probabilities is based on the two basic ingredients  
$\Gamma(t)$ and $v(t)$, which are assumed to depend only on time.
Although this is a widely used simplification, it is not always valid.
As we have mentioned, this approximation is valid for detonations (which are supersonic) 
or for 
deflagrations with velocities $v\lesssim 0.1$.
In the intermediate case of deflagrations which are not very slow ($0.1\lesssim v\lesssim 0.6$), 
a shock wave moving slightly faster than the speed of sound
reheats the plasma in front of the wall. In this case, the kinematic treatment can 
still be simplified by taking into account the fact that, 
in a region of radius $R_\mathrm{sh}\approx (c_s/v)R$ around each bubble, 
the nucleation rate vanishes \cite{lms12}. This approximation assumes that the reheating caused by a single bubble is enough to turn off the nucleation rate, so the inhomogeneous temperature resulting from several shock waves is irrelevant. 
However, some approximation is still required for the velocity, which is not as sensitive to temperature. Our results can in principle be adapted to such a treatment.

Some generalizations of our derivations are straightforward, such as
considering more than two points on bubble walls and calculating 
the probability that some of them are collided and some others are not. 
However, further development would be necessary in order to 
address certain applications such as 
the study of percolation or a thorough calculation topological defect formation, 
which require considering the dynamics of multiple bubble collisions.
Here we have focused on two-point probabilities, and we have used these 
probabilities to discuss
surface correlations between different bubbles, between a single bubble at different times, 
and between different parts of a bubble wall at a given time.
These quantities are relevant for the consequences of the transition, and we have discussed 
a few examples.

In the first place, we have discussed the importance of the total uncollided wall area
$\langle S_\mathrm{tot}\rangle$
as a weight for baryogenesis in an electroweak phase transition with a varying wall velocity.
Indeed, a small velocity is favorable for baryogenesis, 
and in this case the velocity
varies due to reheating during the transition.
We have considered adequate analytic approximations for the wall velocity for this case, 
and we have seen that, depending on the model,
the uncollided wall area will amplify the possible enhancement or suppression 
of the baryon asymmetry due to the velocity variation.

We have argued that the uncollided wall area plays a relevant role in 
the generation of gravitational waves from bubble walls.
This includes, e.g., GW generation from corrugation instabilities, but is 
especially important for the bubble collision mechanism. 
We have analyzed the role of the quantity $\langle S(t)\rangle$ as a time window 
for the source of gravitational waves 
and the dependence of the GW spectrum on
the correlation function $\langle S(t)S(t')\rangle$. In particular, we have shown that the 
latter, without further considerations on the spatial dependence of the source,  
reproduces the correct shape of the GW spectrum, the peak frequency, and the asymptotic behavior at low and high frequencies. 
We have seen that the qualitative behavior of surface correlations is similar
for two very different models of 
bubble nucleation, namely, a delta-function rate and an exponential rate. 
Therefore, we expect the general characteristics of the spectrum to be similar
for different models.
This conclusion is supported by specific calculations \cite{jlst17}.

In contrast, we
have argued that the spatial correlation on a bubble wall is particularly relevant for 
topological defect formation, and this correlation may behave very 
differently depending on the characteristics of bubble nucleation. 
For instance, for an exponential nucleation rate, 
we have a continuous nucleation of very small bubbles. These small bubbles nucleate 
in the false vacuum regions between larger bubbles and are captured 
by the latter before they can collide with each other. 
In contrast, if all the bubbles nucleate simultaneously,
the collisions occur between bubbles of the same size, and we expect that 
in this case it will be easier to enclose a false vacuum region.
We have reinforced this argument by analyzing
the probability for two points on a bubble wall to be collided.

\section*{Acknowledgments}

This work was supported by CONICET grant PIP 11220130100172 
and Universidad Nacional de Mar del Plata, grant EXA897/18.

\appendix

\section{Bubble configurations for two point conditional probability}

In this appendix we consider the two orderings of the times $t_{N},t_{N}^{\prime},t'$
which were not considered in Fig.~\ref{figcorr2b}.

\begin{figure}[tbh]
\centering
\includegraphics[width=1\textwidth]{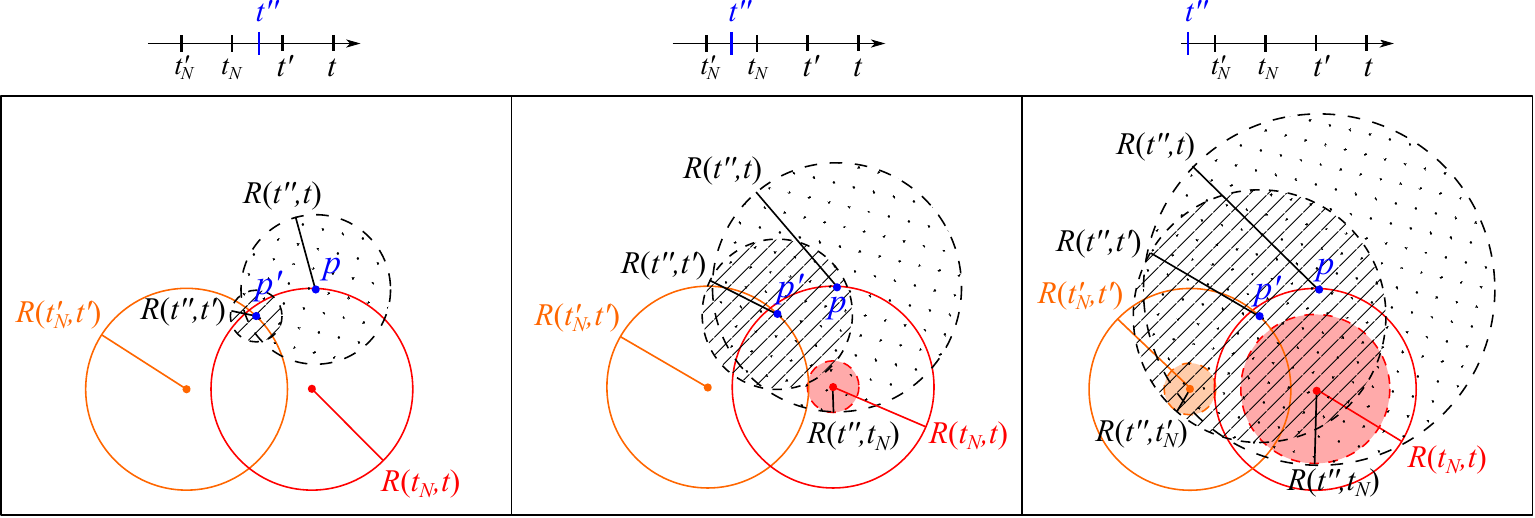}
\caption{Like Fig.~\ref{figcorr2b}, for the time ordering $t_{N}^{\prime}<t_{N}<t'$.
\label{figcorr2b2}}
\end{figure}

The case $t_{N}^{\prime}<t_{N}<t'$ is illustrated in Fig.~\ref{figcorr2b2} for some values of $t''$.
The configuration for  $t''>t'$ is not shown, since in this case the nucleation
at $t''$ cannot affect the events at times $t'$, $t_{N}^{\prime}$
or $t_{N}$, and we have 
\begin{equation}
dP(t'')=dt''\Gamma(t'')\frac{4\pi}{3}R(t'',t)^{3}\qquad(t''>t').
\end{equation}
 In the case $t_{N}<t''<t'$ (left panel), a bubble nucleated at time
$t''$ may have eaten the point $p'$ at time $t'$, so we must exclude
the striped region, 
\begin{equation}
dP(t'')=dt''\Gamma(t'')\left[\frac{4\pi}{3}R(t'',t)^{3}-V_{\cap}\right]\qquad(t_{N}^{\prime}<t''<t').\label{dPapen0}
\end{equation}
For $t_{N}^{\prime}<t''<t_{N}$ (central panel), the nucleation at
$t''$ may also prevent the nucleation of bubble $B$. Therefore,
we must exclude the pink region as well as the striped region, and we
have 
\begin{equation}
dP(t'')=dt''\Gamma(t'')\left[\frac{4\pi}{3}R(t'',t)^{3}-\frac{4\pi}{3}R(t'',t_{N})^{3}-V_{\cap}+V_{\cap}^{\prime}\right]\quad (t_{N}^{\prime}<t''<t_{N}).\label{dPapen}
\end{equation}
 Finally, for $t''<t_{N}^{\prime}$, (right panel), the nucleation
at $t''$ may also prevent the nucleation of bubble $B'$, but the
region which can affect this event (orange shade) is completely contained
within the striped region, and we obtain again Eq.~(\ref{dPapen}).

\begin{figure}[tbh]
\centering
\includegraphics[width=1\textwidth]{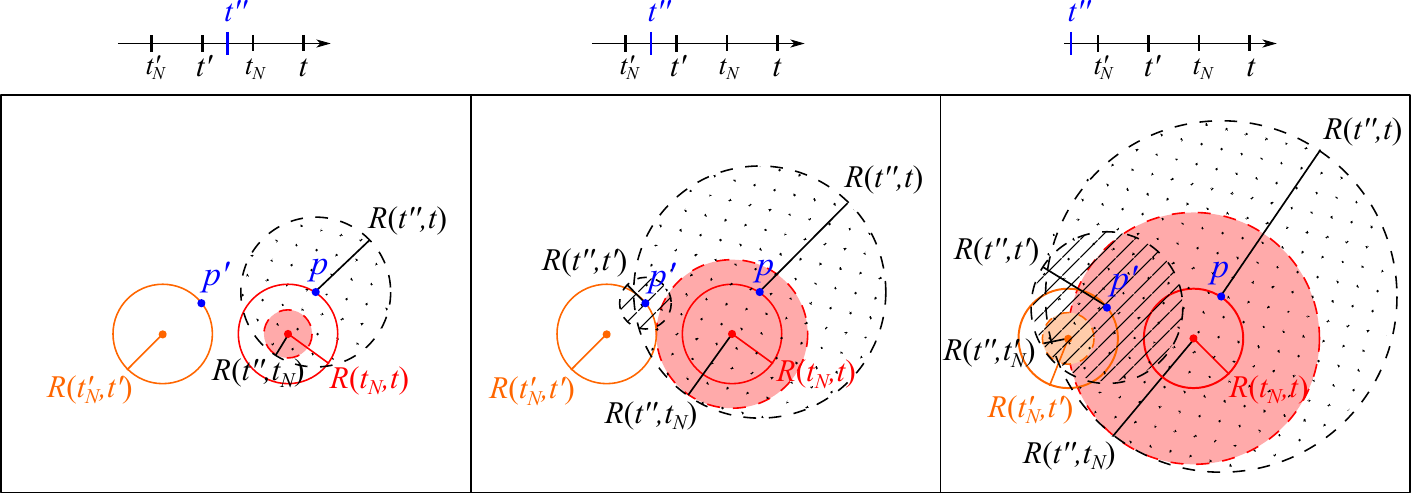}
\caption{Like Fig.~\ref{figcorr2b}, for the time ordering $t_{N}^{\prime}<t'<t_{N}$. 
\label{figcorr2b3}}
\end{figure}

Now let us consider the case $\ensuremath{t_{N}^{\prime}<t'<t_{N}}$, which is illustrated in Fig.~\ref{figcorr2b3}.
For $t''>t'$, the bubble $B'$ and the point $p'$ cannot
be affected by nucleations at $t''$, and we have, for the case $t''>t_N$ (not shown in the figure)
\begin{equation}
dP(t'')=dt''\Gamma(t'')\frac{4\pi}{3}R(t'',t)^{3}\qquad (t''>t_{N}),
\end{equation}
while for the case $t''<t_N$ (left panel),
\begin{equation}
dP(t'')=dt''\Gamma(t'')\frac{4\pi}{3}\left[R(t'',t)^{3}-R(t'',t_{N})^{3}\right]\qquad (t'<t''<t_{N}).
\end{equation}
For $t_{N}^{\prime}<t''<t'$ (central panel), the point $p'$ can be affected, and
we have 
\begin{equation}
dP(t'')=dt''\Gamma(t'')\left[\frac{4\pi}{3}R(t'',t)^{3}-\frac{4\pi}{3}R(t'',t_{N})^{3}-V_{\cap}+V_{\cap}^{\prime}\right]
\quad (t_{N}^{\prime}<t''<t').\label{dPapen2}
\end{equation}
Finally, for $t''<t_{N}^{\prime}$ (right panel), the nucleation of $B'$ can also
be affected, but this is already taken into account in Eq.~(\ref{dPapen2}),
since the orange region is always contained in the striped region.

These results lead to Eqs.~(\ref{Pcond2}-\ref{Iintpri}). In particular, Eqs.~(\ref{dPapen0})
and (\ref{dPapen2}) show that the volume $V_{\cap}$ appears for
$t''<t'$, as expressed by the upper limit of the integral (\ref{Iint}), while Eqs.~(\ref{dPapen}) and (\ref{dPapen2}) show that
the volume $V_{\cap}^{\prime}$ appears for $t''<\min\{t_{N},t'\}$, as expressed by the upper limit of the integral (\ref{Iintpri}).


\end{document}